\documentclass[secnumarabic,superscriptaddress,amssymb,aps,twocolumn]{revtex4-2}

\usepackage{pdfpages} 
\usepackage{pgffor} 

\makeatletter
\AtBeginDocument{\let\LS@rot\@undefined}
\makeatother

\newif\ifarXiv
\arXivfalse

\usepackage{amsmath}
\usepackage{mathrsfs}  
\usepackage{mathtools} 
\usepackage{bbold}   
\usepackage{braket}    
\usepackage{graphicx}				
\usepackage{amssymb}
\usepackage{gensymb} 
\usepackage{wrapfig}
\usepackage{multirow}    
\usepackage{here} 		
\usepackage{url}		
\usepackage[plainpages=false,pdfpagelabels=true,    linkcolor=cyan,breaklinks,colorlinks=true]{hyperref}  
\usepackage{soul}               
\usepackage[dvipsnames]{xcolor}	
\usepackage[multiple]{footmisc}

\usepackage{xspace}
\usepackage{booktabs}  
\usepackage{newunicodechar}
\usepackage{color}
\usepackage{dcolumn}
\usepackage{bm}
\usepackage{mhchem}
\usepackage{verbatim}
\usepackage{tikz}
\usepackage{float}

\definecolor{darkblue}{rgb}{0,0.02,0.45}
\definecolor{darkred}{rgb}{0.45,0.02,0}

\def\be{\begin{equation}}
\def\ee{\end{equation}}
\def\bea{\begin{eqnarray}}
\def\eea{\end{eqnarray}}

\def\vec{\mathbf}

\def\mc{\mathcal}

\def\blio{$\beta$-Li$_2$IrO$_3$\xspace}

\def\hpa{$\textbf{H} \parallel \textbf{a}$\xspace}
\def\hpb{$\textbf{H} \parallel \textbf{b}$\xspace}
\def\hpc{$\textbf{H} \parallel \textbf{c}$\xspace}

\begin{document}

\title{Eclipsing Kitaev: off-diagonal exchange governs 
the correlated high-field phases of $\beta$-Li$_2$IrO$_3$}

\author{Vikram Nagarajan}
\email{vikram.nagarajan@ista.ac.at}
\affiliation{Institute of Science and Technology Austria, 
    3400 Klosterneuburg, Austria}
\author{Ioannis Rousochatzakis}
\affiliation{Department of Physics, Loughborough University, 
    Loughborough LE11 3TU, United Kingdom}
\author{Yuanqi Lyu}
\affiliation{Department of Physics, University of California, 
    Berkeley, California 94720, USA}
\affiliation{Materials Science Division, Lawrence Berkeley 
    National Laboratory, Berkeley, California 94720, USA}
\author{Darian Hall}
\affiliation{Department of Physics, University of California, 
    Berkeley, California 94720, USA}
\affiliation{Materials Science Division, Lawrence Berkeley 
    National Laboratory, Berkeley, California 94720, USA}
\author{Augusto Ghiotto}
\affiliation{Department of Physics, University of California, 
    Berkeley, California 94720, USA}
\affiliation{Materials Science Division, Lawrence Berkeley 
    National Laboratory, Berkeley, California 94720, USA}
\author{Josue Rodriguez}
\affiliation{Department of Physics, University of California, 
    Berkeley, California 94720, USA}
\affiliation{Materials Science Division, Lawrence Berkeley 
    National Laboratory, Berkeley, California 94720, USA}
\author{Koh Yamakawa}
\affiliation{Department of Physics, University of California, 
    Berkeley, California 94720, USA}
\affiliation{Materials Science Division, Lawrence Berkeley 
    National Laboratory, Berkeley, California 94720, USA}
\author{James Analytis}
\affiliation{Department of Physics, University of California, 
    Berkeley, California 94720, USA}
\affiliation{Materials Science Division, Lawrence Berkeley 
    National Laboratory, Berkeley, California 94720, USA}
\author{John Singleton}
\affiliation{National High Magnetic Field Laboratory, 
    Los Alamos National Laboratory, Los Alamos, 
    New Mexico 87545, USA}
\author{Mun K. Chan}
\affiliation{National High Magnetic Field Laboratory, 
    Los Alamos National Laboratory, Los Alamos, 
    New Mexico 87545, USA}
\author{Natalia B. Perkins}
\affiliation{School of Physics and Astronomy, University 
    of Minnesota, Minneapolis, Minnesota 55455, USA}
\author{K.~A. Modic}
\affiliation{Institute of Science and Technology Austria, 
    3400 Klosterneuburg, Austria}

\date{\today}

\begin{abstract}
We report a high-field thermodynamic study of the hyperhoneycomb Kitaev material $\beta$-Li$_2$IrO$_3$, using magnetotropic susceptibility to resolve its low-temperature field-angle phase diagram across the principal crystallographic planes in magnetic fields up to $60$ T. Rather than evolving directly from the low-field incommensurate state into a polarized regime, the system exhibits a strongly direction-dependent sequence of correlated phases. Most notably, for fields in the $ac$-plane, we identify an additional high-field phase that is absent in the other principal planes and exists only within a restricted region of field-angle space. This phase structure is naturally explained by the competition between magnetic field and bond-directional exchange interactions. Using a symmetry-based description supported by microscopic calculations within the $J$-$K$-$\Gamma$ model, we show that off-diagonal $\Gamma$ exchange couples the ferromagnetic and staggered magnetic orders and thereby stabilizes the observed correlated high-field phases. The measured angular dependence of the critical fields is quantitatively captured by this theory, identifying $\Gamma$ exchange as the key interaction controlling the high-field response. These results clarify why the promise of a field-induced spin liquid---the notion that suppressing magnetic order might reveal the underlying Kitaev physics---remains unfulfilled in candidate materials: even when the Kitaev interaction is large, off-diagonal exchange stabilizes symmetry-constrained correlated phases that instead preempt the polarized state.
\end{abstract}

\maketitle

\section*{Introduction}

\begin{figure}
\centering
\includegraphics[width=.9\columnwidth]{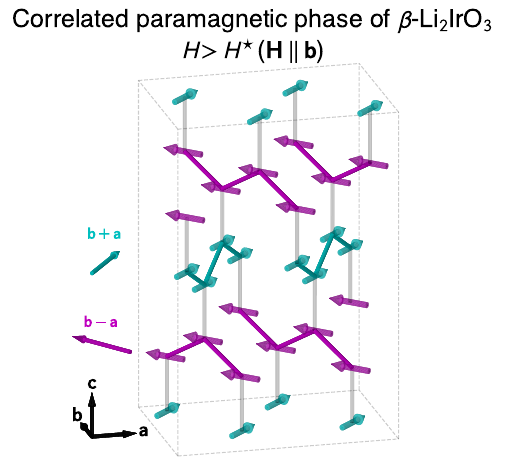}
\caption{The hyperhoneycomb structure of \blio with the 
orthorhombic unit cell doubled along the $a$-axis 
(2$a$,$b$,$c$) 
to highlight the two zigzag chain 
families of the field-induced phase above $H^\star$. 
Cyan and magenta bonds indicate the zigzag chains which propagate along the the $\textbf{a}+\textbf{b}$ and $\textbf{a}-\textbf{b}$ directions, respectively. These chains form the basis of the two-sublattice order that exists above $H^\star$. The precise canting direction of the spins depends on the specific field orientation; for \hpb just above $H^\star$, the spins point nearly parallel (along $\textbf{b}+\textbf{a}$) or antiparallel (along $\textbf{b}-\textbf{a}$) to the zigzag chains, as indicated above.}
\label{fig:Figure1}
\end{figure}

Bond-directional exchange interactions give rise to highly unconventional magnetic-field responses \cite{Janssen2016,Winter2018PRL,Chern2017,Janssen2019,ducatman-2018-magnet-struc, rousochatzakis-2018-magnet-field,li-2020-uncon-magnet,li-2020-reent-incom} in Kitaev materials~\cite{hermanns-2018-physic-kitaev-model,takagi-2019-concep-realiz,trebst-2022-kitaev-mater,Rousochatzakis2024,Tsirlin2022,MatsudaRMP2025}. In several candidates, an external magnetic field suppresses low-temperature order without simply driving the system into a polarized paramagnet. Instead, the field can stabilize nontrivial field-induced phases \cite{BalzPRB2019,ruiz-2017-correl-states,ruiz-2020-high-temper,modic-2017-robus-spin, Choi2019PRB,modic-2020-scale-invar,majumder-2019-anisot-temper, majumder-2020-field-evolut} or, in some cases, a possible spin-liquid regime \cite{banerjee-2018-excit-field,BalzPRB2019,kasahara-2018-major-quant,Bruin2021,Czajka2021,yokoi-2021-half-integ, ruiz-2021-magnon-spinon}. This evolution is often strongly direction-dependent, reflecting the underlying bond-directional character of the exchange interactions.

\begin{figure*}
\centering
\includegraphics[]{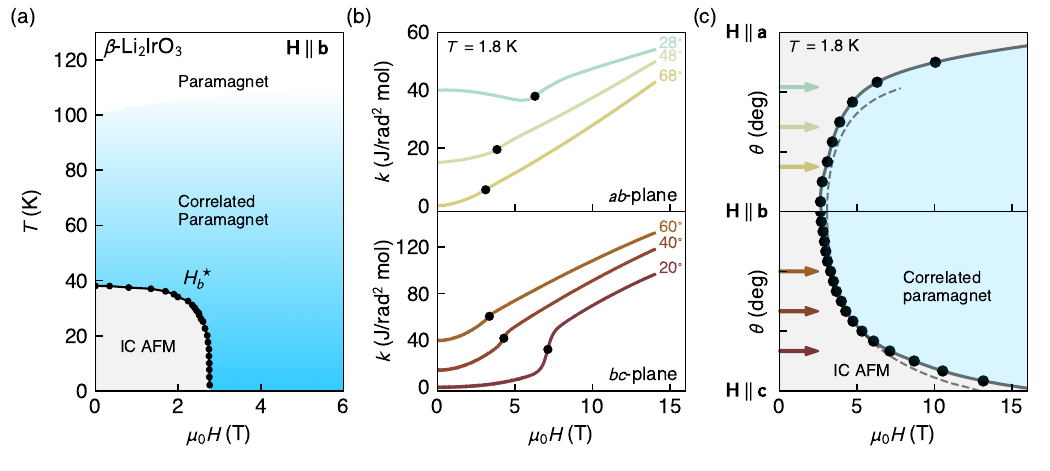}
\caption{(a) Experimentally-determined phase diagram of \blio for \hpb, adapted from \cite{ruiz-2017-correl-states, ruiz-2020-high-temper}. At zero field, \blio enters an incommensurate antiferromagnetic (IC AFM) state below $T_\text{N} = 38$~K. For \hpb at the lowest temperatures, the IC state is suppressed at $H^\star \approx 2.8$~T, giving way to a correlated paramagnetic phase above $H^\star$~\cite{ruiz-2017-correl-states,rousochatzakis-2018-magnet-field}. Additionally, near 100 K, ferromagnetic correlations develop (indicated with blue shading) as measured by thermodynamics and $\mu$SR \cite{ruiz-2020-high-temper}. (b) Magnetotropic susceptibility for the $ab$- and $bc$-planes of \blio at $T = 1.8$~K up to 14~T. As the field is rotated away from the $b$-axis, the IC transition field shifts to higher magnetic fields. $H^\star$ transitions are indicated by the solid black dots. Curves are offset vertically for clarity. (c) The IC AFM phase boundary for the $ab$- and $bc$-planes measured up to 14~T, with $H^\star$ indicated by black dots connected with solid grey lines to guide the eye. The colored arrows at different angles correspond to the representative curves shown in panel (b). Dashed lines indicate the theoretical IC phase boundary, as determined by classical energy minimization for the parameter set $(J,K,\Gamma)=(0.4,-18,-10)$ meV and an isotropic $g$-factor with $g=2$.} 
\label{fig:Figure2}
\end{figure*}

The hyperhoneycomb iridate \blio (\autoref{fig:Figure1}), described by a dominant Kitaev interaction, provides a particularly important platform for this problem \cite{biffin-2014-uncon-magnet, takayama-2015-hyper-iridat,Tsirlin2022}. Importantly, despite the dominant Kitaev exchange $K$, additional off-diagonal exchange $\Gamma$ and Heisenberg exchange $J$ stabilize a low-temperature ordered phase, precluding a quantum spin liquid ground state. Below $T_\text{N} \approx 38$~K, \blio exhibits an incommensurate (IC) counter-rotating spiral magnetic order with propagation vector $\mathbf{Q}\approx(0.57,0,0)$~\cite{biffin-2014-uncon-magnet}. Previous theoretical studies have utilized the minimal $J$-$K$-$\Gamma$ model with an additional term for the magnetic field
\begin{equation}
    \mc{H}=\sum_t\sum_{\langle ij\rangle\in t}
    \mc{H}_{ij}^t+\mc{H}_{Z},
\end{equation}
where $\mc{H}_{ij}^t$ is the nearest-neighbor 
$J$-$K$-$\Gamma$ Hamiltonian
\begin{equation}\label{eq:jkg}
    \mc{H}_{ij}^t=J\vec{S}_i\cdot\vec{S}_j
    +KS_i^{\alpha_t}S_j^{\alpha_t}
    +\sigma_t\Gamma(S_i^{\beta_t}S_j^{\gamma_t}
    +S_i^{\gamma_t}S_j^{\beta_t}),
\end{equation}
and $\mc{H}_Z$ is the Zeeman term
\begin{equation}\label{eq:HZ1}
    \mc{H}_{Z}=-\mu_B\vec{H}\cdot\sum_{i}\vec{g}_i
    \cdot\vec{S}_i.
\end{equation}
These studies predicted an unusually anisotropic field-angle phase diagram, including the suppression of IC order and the emergence of symmetry-constrained high-field states~\cite{ducatman-2018-magnet-struc,rousochatzakis-2018-magnet-field,li-2020-reent-incom, li-2020-uncon-magnet}. Experimentally, however, only the field response for $H \parallel b$ has been established, where the IC order collapses at $H^\star \approx 2.8$~T (\autoref{fig:Figure2}a). Beyond $H^\star$, the field-induced state is described by coexisting ferromagnetic and zigzag components~\cite{ruiz-2017-correl-states,ruiz-2021-magnon-spinon}, which is illustrated in \autoref{fig:Figure1}. The broader field-angle phase diagram and the nature of the high-field phases remains relatively unexplored.

Here, we use magnetotropic susceptibility measurements to map the field-angle phase diagram of \blio across the principal crystallographic planes in magnetic fields up to $60$~T. We uncover a strongly direction-dependent sequence of correlated phases and identify an additional high-field phase for fields in the $ac$-plane that is absent for other field orientations. Supported by a symmetry-based theoretical analysis, our results show that off-diagonal $\Gamma$ exchange couples ferromagnetic and staggered magnetic order parameters and thereby controls the high-field response of \blio. 

\section*{Experimental Results}

We probe the high-field magnetic response of \blio using magnetotropic susceptibility as measured by resonant torsion magnetometry (RTM)~\cite{modic-2018-reson-torsion,shekhter-2023-magnet-suscep}. The magnetotropic susceptibility, $k = \partial^2 F/\partial \theta^2 $, is the second-derivative of the magnetic free energy with respect to magnetic field orientation, i.e., the angular derivative of the magnetic torque. As such, it is particularly sensitive to anisotropic phase transitions, providing novel insight into unconventional superconductors \cite{Zambra2026} and frustrated quantum magnets \cite{Gui2025, Nasir2026}. By rotating magnetic field within the three crystallographic planes, this technique enables a systematic mapping of the field-angle phase diagram in high magnetic fields \footnotemark[1].

\autoref{fig:Figure2}b shows the magnetotropic susceptibility in the $ab$- and $bc$-planes up to 14~T. For magnetic field applied along the $b$-axis in both planes, the IC order is suppressed at $H^\star = 2.8$~T, consistent with previous studies~\cite{ruiz-2017-correl-states, li-2020-uncon-magnet}. As the field is rotated away from the $b$-axis, the critical field increases rapidly, demonstrating the strong anisotropy of the IC phase in both the $ab$- and $bc$-planes (\autoref{fig:Figure2}c).

Extending these measurements in the $ab$- and $bc$-planes to pulsed magnetic fields reveals an even steeper increase of the critical field needed to suppress IC order as field approaches the $a$- and $c$-axes, as shown in \autoref{fig:Figure3}a and Supplemental Material (SM) Figure S3~\footnote[1]{\label{SM}See Supplemental Material at [URL will be inserted by publisher] for additional details on sample preparation, resonant torsion magnetometry, pulsed-field data in the $bc$-plane, symmetry analysis, and free energy minimization calculations, which includes Refs.~\cite{lee-2015-theor-magnet,lee-2016-two-iridat,riedl-2019-sawtoot-torque,modic-2014-realiz-three}}. For all field orientations in the $ab$- and $bc$-planes, no additional phase transitions are observed beyond $H^\star$, and the system evolves continuously toward the field-polarized regime (\autoref{fig:Figure3}a and SM Figures~S2 and~S3 \footnotemark[1]).

In contrast to the $ab$- and $bc$-planes, measurements in the $ac$-plane reveal qualitatively new behavior (\autoref{fig:Figure3}b). While the IC phase is suppressed at $H^\star$, we observe an additional phase transition at higher fields, which we denote with $H^{\star\star}$. This transition evolves from a small change in slope for \hpc to a prominent feature in $k$ at intermediate field angles, which is consistent with the curvature of the phase boundary and indicative of a distinct thermodynamic phase~\cite{modic-2018-reson-torsion,shekhter-2023-magnet-suscep, modic-2020-scale-invar}. 

In the $ac$-plane, the lower critical field $H^\star$ increases from $\sim$16~T for \hpc to beyond the accessible field range for \hpa. The higher field transition $H^{\star\star}$ follows the same trend, occurring at $\sim$42~T for \hpc, and increasing rapidly upon rotation toward the $a$-axis.

\begin{figure*}[ht!]
\centering
\includegraphics[width=0.8\textwidth]{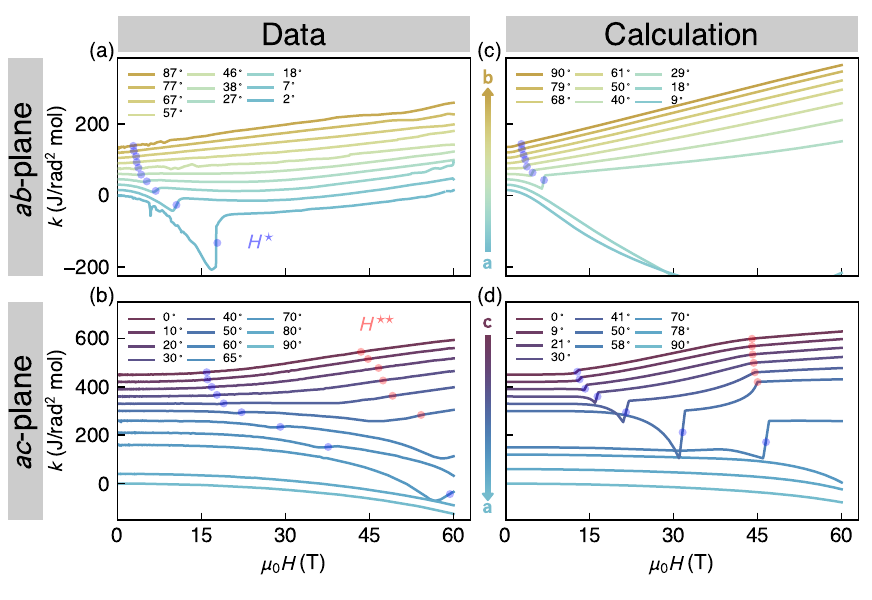}
\caption{Magnetotropic susceptibility for the $ab$- and $ac$-planes of \blio at $T \sim 1.5$--$4$~K in pulsed magnetic fields up to 60~T (curves are offset for clarity). (a) Representative field orientations in the $ab$-plane. The IC phase transition at $H^\star$ appears as a drop in the magnetotropic susceptibility below 20 T for all field orientations. (b) The magnetotropic susceptibility up to 60 T for fixed field orientations in the $ac$-plane. In addition to the IC phase transition at $H^\star$, a second phase transition at $H^{\star\star}$ is observed at higher fields. Both features evolve to be more prominent as field rotates from \hpc ($\theta$ = 0$^\circ$) to \hpa ($\theta$ = 90$^\circ$). Within 25$^\circ$ of the $a$-axis, no transition is observed. (c,d) Calculated magnetotropic susceptibility for similar field orientations as (a) and (b), respectively, using ($J$,$K$,$\Gamma$) = (0.4, -18, -10) meV, and $g_{aa}$=$g_{bb}$=$g_{cc}$=2, $g_{ab}$=0.1---values taken from \citet{li-2020-reent-incom}. 
For \hpa in the $ac$-plane, the IC phase transition is beyond the measurable field range, in agreement with theoretical predictions.
See \autoref{fig:FigureEM2} and \autoref{fig:FigureEM1} in the End Matter for details on how the transition fields were determined.
}
\label{fig:Figure3}
\end{figure*}

\section*{Symmetry Origin of the High-Field Phase}
\begin{figure*}[ht!]
\centering
\includegraphics[]{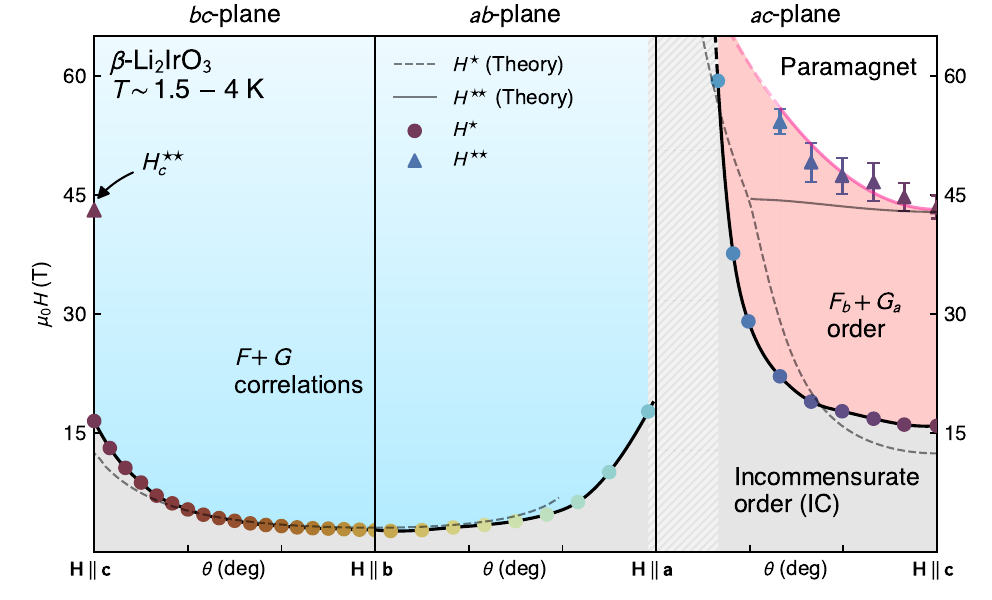}
\caption[]{Experimental low-temperature and high-field phase diagram for \blio determined by magnetotropic susceptibility measurements. Solid circles represent the IC phase boundary ($H^\star$), and solid triangles represent the transition into the field-polarized phase ($H^{\star\star}$). The grey shaded region represents the IC phase, while the blue region within the $ab$- and $bc$- planes indicates strong $F+G$ correlations, where the two-sublattice magnetic state exists without spontaneous symmetry breaking---the relevant symmetries are already broken explicitly by the applied field. The red region in the $ac$-plane indicates a distinct ordered phase in which $F_b$ and $G_a$ order parameters develop spontaneously, breaking $\Theta C_{2\textbf{b}}$. This spontaneous symmetry breaking occurs exclusively within the $ac$-plane: a magnetic field oriented within this plane preserves $\Theta C_{2\textbf{b}}$ as the only remaining point-group symmetry, which is broken due to the $\Gamma$-mediated cross-coupling above $H^\star$ (SM Section S4). Dashed lines indicate the theoretical IC phase boundary,  as determined by classical energy minimizations for the refined parameter set $(J,K,\Gamma)=(0.4,-24,-9.3)$ meV. The theoretical paramagnetic phase boundary is also indicated by the solid grey line in the $ac$-plane.
Solid black and pink lines are guides to the eye, with dashed extensions indicating regions beyond experimentally accessible magnetic fields. The light grey region near \hpa also indicates the angle range where transition fields are too high to be observed.}
\label{fig:Figure4}
\end{figure*}

Notably, early evidence for a second transition in \mbox{\blio} at $H^{\star \star}$ in the $ac$-plane comes from theoretical calculations of the torque component $\tau_b$ ~\cite{li-2020-reent-incom}. However, the significance of this feature --- namely, that it marks the emergence of a distinct, symmetry-broken phase  that survives away from the $c$-axis --- was not recognized at the time. To understand its origin, we revisit the effective description of the magnetic state above $H^\star$~\cite{li-2020-uncon-magnet}.

Above $H^\star$, the incommensurate component of the multi-sublattice order is absent, leaving a simpler two-sublattice magnetic state. The spins on the $\mathbf{b}+\mathbf{a}$ zigzag chains are described by the sublattice magnetization ${\bf A}$, while those on the $\mathbf{b}-\mathbf{a}$ chains are described by ${\bf A}'$ (\autoref{fig:Figure1}). It is then natural to introduce two uniform (${\bf k}\!=\!0$) symmetry-resolved fields: the net (ferromagnetic) magnetization ${\bf F}=({\bf A}+{\bf A}')/2$ and the transverse staggered (zigzag) component ${\bf G}=({\bf A}'-{\bf A})/2$.

Within the minimal $J$-$K$-$\Gamma$ model proposed for \blio~\cite{ducatman-2018-magnet-struc,rousochatzakis-2018-magnet-field,li-2020-uncon-magnet,li-2020-reent-incom}, the exchange energy contains the following cross-coupling term (the full expressions for the energy are given in SM, Section S3 \footnotemark[1]):
\be\label{eq:GammaFG}
E_{FG} = -\sqrt{2}\Gamma (F_aG_b + F_bG_a)\,.
\ee
For fields in the $ac$-plane, the applied field breaks all point group symmetries of the Hamiltonian except the mirror symmetry, ${\Theta}C_{2\mathbf{b}}$, where $\Theta$ denotes time reversal. The components $F_a$, $F_c$, and $G_b$ are even under ${\Theta}C_{2\mathbf{b}}$, and are continuously induced by the Zeeman field and the first term of \autoref{eq:GammaFG}. In contrast, $F_b$, $G_a$, and $G_c$ are odd under ${\Theta}C_{2\mathbf{b}}$ and can only appear spontaneously. The cross-coupling $\Gamma$ term provides the driving force; below $H^{\star\star}$,
$F_b$ and $G_a$ (and $G_c$, by virtue of the spin-length constraints) become nonzero for a range of field orientations in the $ac$-plane (SM Figure S7 \footnotemark[1]).

Although analogous considerations apply for fields in the $bc$-plane, with $F_a$ and $G_b$ playing the role of potential order parameters, such an instability does not occur. The reason can be traced to the fact that, unlike the $ac$-plane, the IC phase only breaks translations and not the corresponding mirror symmetry, ${\Theta}C_{2a}$. Therefore, an intermediate phase in the $bc$-plane would require two unrelated instabilities to coincide at $H^\star$ for a range of $\theta$. This is not generically expected \footnotemark[2], and such a lack of an intermediate phase for the $bc$-plane is supported by numerics (SM Section S4)~\footnotemark[1]. 
\footnotetext[2]{The possibility for two intermediate phases, where the system first restores translations at $H_{bc}^\ast$, then breaks ${\Theta}C_{2a}$ at $H_{ac}^{**}$, and then restores ${\Theta}C_{2a}$ at $H_{ac}^{***}$ is also not supported by the numerics.}

Finally, for fields along the $ab$-plane, the only components odd under the surviving symmetry ${\Theta}C_{2\mathbf{c}}$ are $F_c$ and $G_c$, neither of which appears in $E_{FG}$, so no coupling mechanism exists and no instability is possible regardless of field strength. Notably, the critical fields $H^\star$ and $H^{\star\star}$ depend on $J$ and $\Gamma$, but are independent of the Kitaev coupling $K$ at leading order~\cite{li-2020-reent-incom}. 
This identifies $\Gamma$ as the interaction governing the field-induced phases in \blio.

\section*{Discussion}

The high-field phase diagram established here is summarized in \autoref{fig:Figure4}, and can contrasted with the theoretical one shown in SM Figure~S5 \footnotemark[1]. The results reveal that the suppression of IC order in \blio does not lead directly to a trivial polarized state. Instead, for fields in the $ac$-plane, the system enters a new high-field phase bounded by a second-order transition $H^{\star\star}$ that appears only in a restricted angular range. This demonstrates that the route to polarization is controlled by the interplay of crystal symmetry and spin-orbit-coupled exchange.

Whether $H^\star$ and $H^{\star\star}$
ultimately merge at a finite field, as suggested by classical numerics for a specific parameter set, or instead persist as separate boundaries extending to higher fields remains an open question. Resolving this would require measurements at higher fields with finer angular resolution near $\theta \approx 70^\circ$ from the
$c$-axis, where the two boundaries approach each other most closely in our data.


The theoretical expectations (SM Figure~S5 \footnotemark[1]) for the angle-dependence of the lower critical field $H^\star$, associated with the collapse of IC order, closely agrees with the experimental values (\autoref{fig:Figure4}). This agreement is particularly striking because the phase boundaries are highly anisotropic, varying from 2.8~T for \hpb to $\sim$16~T for \hpc. The main discrepancy arises for \hpa, where $H^\star$ is expected to approach 100 T  \cite{li-2020-reent-incom}. For fields near the $a$-axis, the largest observed $H^\star$ is $\sim$60 T for measurements performed in the $ac$-plane. In the $ab$-plane, however, $H^\star$ only reaches $\sim$18 T for field aligned at the angle closest to the nominal $a$-axis. A consistent $H^\star$ for \hpa in both measurement planes is required, suggesting that either the field-rotation plane is precessing about the $ab$-plane, or the IC phase boundary rises more steeply than predicted for \hpa in the $ab$-plane (or both). Another discrepancy is the absence of the narrow re-entrant IC region predicted within 30$^\circ$ of the $a$-axis \cite{li-2020-reent-incom}, which is likely difficult to resolve experimentally because it is predicted to occur over a very narrow angular interval close to the edge of the accessible field range.

The observed critical fields also provide a direct determination of the exchange parameters $J$ and $\Gamma$ that is independent of the Kitaev coupling $K$. This is in contrast to spin-wave analyses, where the exchange parameters are constrained by fitting the full magnon dispersion and $J$, $K$, and $\Gamma$
contribute in ways that are difficult to disentangle \cite{Halloran2022, ruiz-2021-magnon-spinon}.   Furthermore, $\Gamma$ is typically constrained by fitting a broad, powder-averaged spectral feature to linear spin-wave theory~\cite{Halloran2022}---an approach that is subject to systematic uncertainties from orientational averaging and the known breakdown of the magnon picture in strongly Kitaev-dominated systems \cite{Winter2018PRL}. The critical fields measured here on single crystals correspond to sharp thermodynamic phase boundaries that place precise constraints on $J$ and $\Gamma$ directly. Using the measured values of $\mu_0H_c^\star \approx 16$~T and $\mu_0H_c^{\star\star} \approx$ 42~T, together with the analytical expressions for the critical fields (SM Section~S4 \footnotemark[1]), we obtain $J \sim 0.5 \pm 0.1$~meV and $\Gamma \sim -11.5 \pm 2.5$~meV.
These values are consistent, within the experimental uncertainty, with those extracted from zero-field neutron spectroscopy ($J = 0.40 \pm 0.02$~meV, $\Gamma = -9.3 \pm 0.1$~meV~\cite{Halloran2022}). The discrepancy likely reflects these distinct systematic sensitivities rather than a failure of the minimal model. Importantly, both determinations place $K$ as the leading contributor and $\Gamma$ as the dominant non-Kitaev interaction.

The persistence of strong correlations above $H^\star$, the symmetry-selective appearance of $H^{\star\star}$, and the angular dependence of the critical fields all follow from $\Gamma$-mediated coupling between the ferromagnetic and zigzag order parameters. The broad intermediate field window between $H^\star$ and full polarization everywhere in the phase diagram demonstrates that suppressing long-range IC order does not eliminate exchange-driven structure; instead, \blio retains a correlated high-field state over a wide range of fields, with its stability and symmetry determined by $\Gamma$. At the highest measured fields in the $bc$-plane (SM Figure~S3 \footnotemark[1]), the crossover toward $g$-factor-dominated anisotropy marks the eventual approach to a conventional spin-polarized regime.


The energy scale set by $\Gamma$ also suggests a connection to the anomalous magnetic behavior observed near 100~K in \blio, well above the ordering temperature $T_\text{N} \approx 38$~K~\cite{ruiz-2020-high-temper, Vranas2026}. The $\Gamma$-mediated coupling between $\mathbf{F}$ and $\mathbf{G}$ sets an energy scale of order $|\Gamma| \sim 10$~meV~$\sim 100$~K, at which short-range $F_b + G_a$ correlations may begin to develop even without long-range order~\cite{li-2020-reent-incom}. The high-field phases identified here, stabilized by precisely this coupling, may therefore have a finite-temperature precursor---a scenario consistent with the anomalous susceptibility and weak (quasi)static ferromagnetic signal reported in prior thermodynamic and $\mu$SR measurements~\cite{ruiz-2020-high-temper,ruiz-2021-magnon-spinon}. Whether the 100~K anomaly reflects such a fluctuation onset---driven by thermal population of $\Gamma$-scale modes rather than field-induced symmetry breaking---remains an open question. 

More broadly, these findings clarify why field-induced spin-liquid behavior is difficult to realize in candidate Kitaev magnets. Even when the Kitaev interaction is large, substantial off-diagonal exchange can reorganize the high-field phase diagram and stabilize symmetry-constrained correlated phases before a trivial polarized state is reached. In this sense, $\Gamma$ is not a perturbative correction to the Kitaev limit, but a defining interaction in the magnetic-field response of strongly spin-orbit-coupled magnets.

\begin{acknowledgments}
The authors thank Brad Ramshaw, Arkady Shekter and Alexander Tsirlin for fruitful discussions. The authors also acknowledge the expert support from the staff scientists at Los Alamos National Laboratory's Pulsed Field Facility in setting up and performing the experiments. The National High Magnetic Field Laboratory is supported by the National Science Foundation through DMR-1644779 and DMR-2128556, the State of Florida, and the U.S. Department of Energy. Measurement and development of magnetotropic susceptibility for pulsed magnetic fields was supported by DOE-BES ``Science of 100T program.'' V.N. and K.A.M. acknowledge funding received from the European Research Council (ERC) under the European Union's Horizon 2020 research and innovation programme (TROPIC-101078696). V.N. acknowledges support from the National Science Foundation Graduate Research Fellowship Grant No. DGE 1752814 and the UC-National Lab In-Residence Graduate Fellowship Grant No. L21GF3660. IR acknowledges the support by the Engineering and Physical Sciences Research Council, Grant No. EP/V038281/1. N.B.P. was supported by the U.S. Department of Energy, Office
of Science, Basic Energy Sciences under Award No. DE-SC0018056.
\end{acknowledgments}


\bibliography{references_final}

\clearpage

\section*{End Matter}

\subsection*{Selecting critical fields}

\begin{figure*}[!h]
  \centering
\includegraphics[]{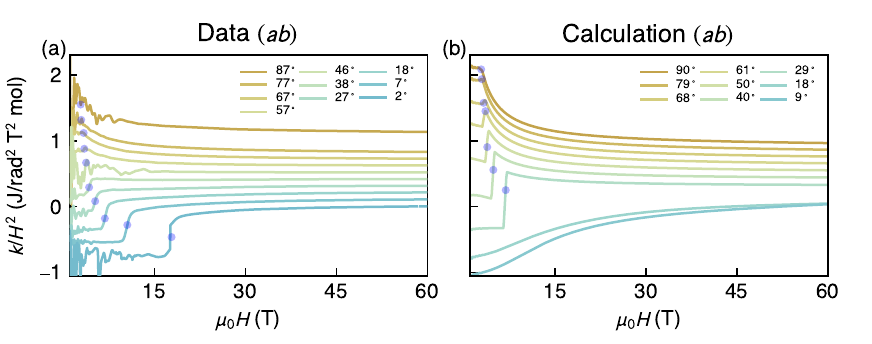}
\caption{Magnetotropic susceptibility divided by magnetic field squared for the $ab$-plane as measured by RTM (a) and calculated (b). Near \hpb, the transition fields can be discerned by noting the shoulder of the curves, denoted with blue circles. These features are consistent with theoretical predictions in panel (b). Away from \hpb, the transitions can more obviously be seen from the jumps in $k$.}
\label{fig:FigureEM2}
\end{figure*}

\begin{figure*}[!h]
  \centering
\includegraphics[]{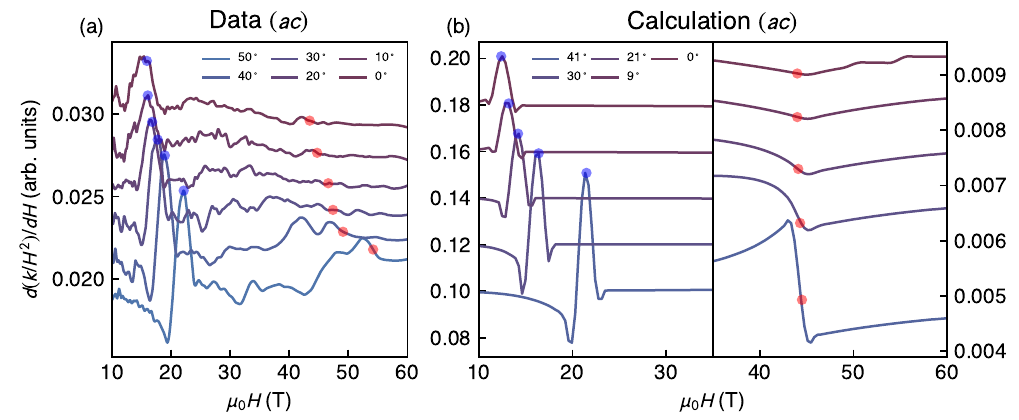}
\caption{Comparison of $d(k/H^2)/dH$ for both experiment (a) and theory (b) for representative angles in the $ac$-plane. In (b), the low-field and high-field regions are shown on separate y-axis scales to better highlight both features. All curves are vertically offset for clarity. $\theta$ is defined as the angle between the applied magnetic field and the $c$-axis. This representation isolates both phase transitions as sharp features in the derivative: $H^\star$ appears as a low-field peak in the derivative (blue dots), marking the collapse of IC order, while $H^{\star\star}$ is identified from the field at which the derivative settles into the flat, featureless behavior of the field-polarized regime (red dots). This behavior is consistent with the classical energy minimization, which predicts that the order parameters of the intermediate phase vanish continuously as $\sqrt{(H^{\star \star}-H)}$ approaching $H^{\star \star}$ from below. This produces a divergent slope in $k$ at the transition and a correspondingly sharp feature in its derivative. The criterion is applied identically to experimental and calculated curves at matching field angles, without independent adjustment. The agreement in field position, angular evolution, and lineshape between experiment and the minimal $J$-$K$-$\Gamma$ theory supports the identification of $H^{\star \star}$ as a genuine phase boundary. 
}
\label{fig:FigureEM1}
\end{figure*}

For continuous phase transitions, the jump in the magnetotropic susceptibility follows the Ehrenfest relation
\begin{align}\label{eq:ehrenfest}
    \Delta k = -\Delta \chi \left( \frac{\partial H_c}{\partial \theta} \right)_T^2.
\end{align}
Because this jump scales with the \textit{square} of the phase boundary slope, transitions become difficult to resolve where $\partial H_c/ \partial \theta \to 0$, which is generically often the case near high-symmetry directions. This difficulty is compounded in pulsed-field measurements, where the short (80 ms) pulse duration limits the achievable signal-to-noise.  

In the $ab$- and $bc$-planes, this only affects the region near \hpb: transitions elsewhere in these planes appear as clear jumps in $k$ (\autoref{fig:Figure3} and SI Figure 3), but near \hpb the transition instead appears as a subtle shoulder. To isolate it, we plot $k/H^2$ (\autoref{fig:FigureEM2}). This removes the leading field dependence and makes the shoulder, which is consistent in field position and shape with the calculated curves, resolvable close $\theta = 0$. 

The $ac$-plane presents a distinct version of the same issue because the jump in $k$ at both $H^\star$ and $H^{\star \star}$ becomes subtle as field is rotated towards the $c$-axis; their phase boundaries flatten for this field-angle and $H^{\star \star}$ is additionally softened because its order parameter grows continuously from zero rather than jumping (see SI Section S5.5). To track both transitions across the full angle range, we compute $d(k/H^2)/dH$ (\autoref{fig:FigureEM1}). Moving away from the $c$-axis, the steepening of both phase boundaries enhances $\Delta k$ through \autoref{eq:ehrenfest} and the transitions become increasingly well-defined at intermediate angles. We identify $H^\star$ from the low-field peak in this derivative, marking the onset of the IC transition. $H^{\star \star}$ is identified from the field at which the derivative turns over and flattens into the field-polarized regime --- the point beyond which no further transitions are resolvable. This criterion is applied consistently across all measured and calculated angles in \autoref{fig:FigureEM1}.

\clearpage

\end{document}


\title{Supplemental Material for ``Eclipsing Kitaev: off-diagonal exchange governs the correlated high-field phases of \blio''}

\author{Vikram Nagarajan}
\email{vikram.nagarajan@ista.ac.at}
\affiliation{Institute of Science and Technology Austria, 
    3400 Klosterneuburg, Austria}
\author{Ioannis Rousochatzakis}
\affiliation{Department of Physics, Loughborough University, 
    Loughborough LE11 3TU, United Kingdom}
\author{Yuanqi Lyu}
\affiliation{Department of Physics, University of California, 
    Berkeley, California 94720, USA}
\affiliation{Materials Science Division, Lawrence Berkeley 
    National Laboratory, Berkeley, California 94720, USA}
\author{Darian Hall}
\affiliation{Department of Physics, University of California, 
    Berkeley, California 94720, USA}
\affiliation{Materials Science Division, Lawrence Berkeley 
    National Laboratory, Berkeley, California 94720, USA}
\author{Augusto Ghiotto}
\affiliation{Department of Physics, University of California, 
    Berkeley, California 94720, USA}
\affiliation{Materials Science Division, Lawrence Berkeley 
    National Laboratory, Berkeley, California 94720, USA}
\author{Josue Rodriguez}
\affiliation{Department of Physics, University of California, 
    Berkeley, California 94720, USA}
\affiliation{Materials Science Division, Lawrence Berkeley 
    National Laboratory, Berkeley, California 94720, USA}
\author{Koh Yamakawa}
\affiliation{Department of Physics, University of California, 
    Berkeley, California 94720, USA}
\affiliation{Materials Science Division, Lawrence Berkeley 
    National Laboratory, Berkeley, California 94720, USA}
\author{James Analytis}
\affiliation{Department of Physics, University of California, 
    Berkeley, California 94720, USA}
\affiliation{Materials Science Division, Lawrence Berkeley 
    National Laboratory, Berkeley, California 94720, USA}
\author{John Singleton}
\affiliation{National High Magnetic Field Laboratory, 
    Los Alamos National Laboratory, Los Alamos, 
    New Mexico 87545, USA}
\author{Mun K. Chan}
\affiliation{National High Magnetic Field Laboratory, 
    Los Alamos National Laboratory, Los Alamos, 
    New Mexico 87545, USA}
\author{Natalia B. Perkins}
\affiliation{School of Physics and Astronomy, University 
    of Minnesota, Minneapolis, Minnesota 55455, USA}
\author{K.~A. Modic}
\affiliation{Institute of Science and Technology Austria, 
    3400 Klosterneuburg, Austria}

\date{\today}

\maketitle


\section{\label{sec:setup}Experimental setup}

\subsection{Crystal synthesis and morphology}
Single crystals of \blio{} were synthesized via the standard self-vapor transport technique \cite{ruiz-2020-high-temper,ruiz-2021-magnon-spinon}. Crystals grow in a rhombus shape, with the $a$- and $b$-axes pointing along the short and long diagonals, respectively, and the $c$-axis pointing perpendicular to the face of the rhombus (the $ab$-plane). A single crystal mounted for $ac$-plane measurements is shown in \cref{fig:figure1}.

\subsection{Magnetotropic susceptibility}


\begin{figure}[ht!]
  \centering
\includegraphics[width=0.8\columnwidth,trim=0cm 0cm 0cm 0cm, clip=true]{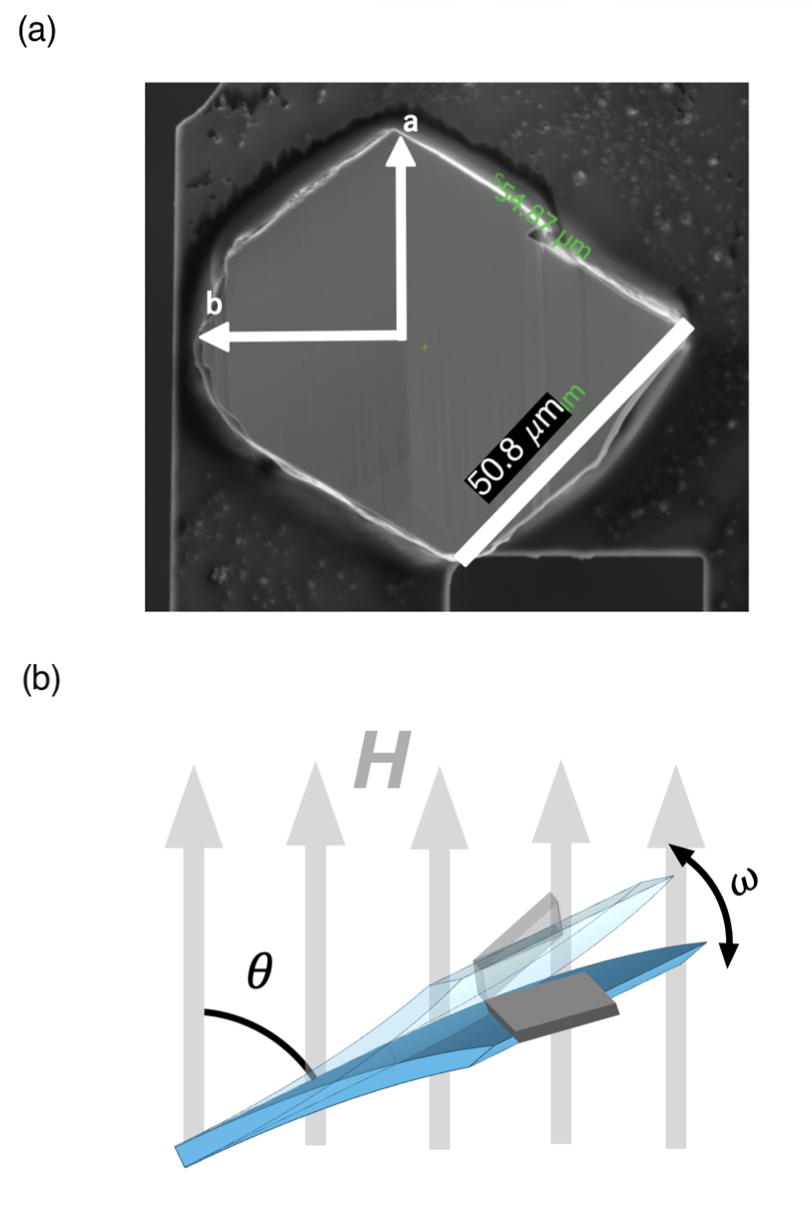}
  \caption{a) An image of a single crystal of \blio mounted for 
  resonant torsion experiments for magnetic field applied in the $ac$-plane (\textit{i.e.} the $a$-axis is aligned with the long axis of the lever). The crystallographic $a$- and $b$-axes are indicated with the white arrows. The thickness of the sample is $\sim$10 microns. b) Schematic of the cantilever vibrating at $\sim$40 kHz with respect to the external magnetic field. $\theta$ is always defined as the angle between the external field and one of the crystallographic axes.}
  \label{fig:figure1}
\end{figure}

The magnetotropic susceptibility, \(k\), is the derivative of the magnetic torque with respect to field angle, and is therefore a second derivative of the magnetic free energy
\begin{equation} \label{eq:k-def}
    k = \frac{\partial \tau}{\partial \theta} = \frac{\partial^2 F_\textnormal{mag} }{\partial \theta^2}.
\end{equation}
As a second derivative of the free energy, \(k\) is particularly sensitive to anisotropic magnetic phase transitions and provides a powerful probe of the field-angle phase diagram. Furthermore, the magnetotropic response is largest near high-symmetry crystallographic directions, where the magnetic torque itself vanishes. As a result, magnetotropic susceptibility 
allows for the robust characterization of magnetic anisotropy with field applied along arbitrary directions. We utilize these properties for our comprehensive characterization of the magnetic response for \blio along all principal crystallographic axes and planes.

The magnetotropic susceptibility is detected by resonant torsion magnetometry (RTM) \cite{modic-2018-reson-torsion,shekhter-2023-magnet-suscep}. The sample is placed on a vibrating cantilever with a bare resonant frequency, \(f_0\). The second angular derivative of the magnetic free energy of the sample is measured as a shift in the resonant frequency of the cantilever as the sample-cantilever system oscillates in an external magnetic field. The change in the resonant frequency \(\Delta f\) with applied magnetic field \(H\) is proportional to \(k\) (assuming $\Delta f \ll f_{0}$), given by the relation
\begin{equation}\label{eq:mt-freq}
  \frac{\Delta f(\theta, H)}{f_0}=\frac{k(\theta, H)}{2 S},
\end{equation}
where \(\theta\) is the magnetic field orientation and \(S\) is the bare cantilever stiffness \cite{modic-2018-reson-torsion, shekhter-2023-magnet-suscep}. Therefore, if the resonant frequency and effective bending stiffness of the cantilever is known for \(H = 0\), the magnetotropic susceptibility can be determined directly from the frequency shift. Alternatively, one can use the angle-dependence of $k$ in the linear response regime, combined with known magnetic susceptibility anisotropy, to extract $k$ from $\Delta f$ (see \cref{subsec:conversion}). 

\begin{figure}[ht!]
\includegraphics[]{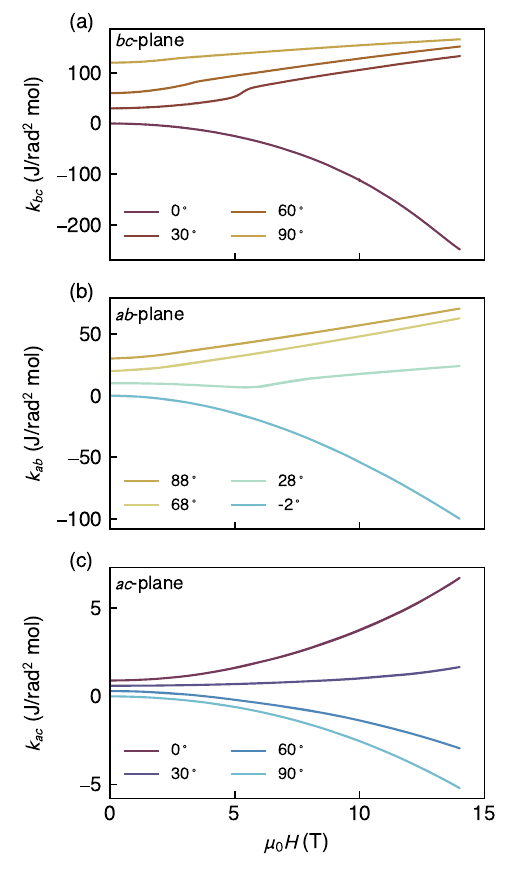}
\caption{
Magnetotropic susceptibility $k$ as a function of magnetic field up to 14 T for the three principal crystallographic planes. Different curves correspond to different field orientations within each plane, with vertical offsets for clarity. The angle $\theta$ is measured from the $c$-axis in the $bc$-plane (a) and $ac$-plane (c), and from the $a$-axis in the $ab$-plane (b). The suppression of the IC phase at $H^\star$ evolves from 2.8 T for \hpb to beyond the measured field range of 14 T for magnetic field applied near the $a$- and $c$-axes. This results in a quadratic field-dependence of $k$ for field orientations where $H^\star$ exists beyond the measured field range.}
\label{fig:figure2}
\end{figure}

The magnetotropic susceptibility was measured both in DC magnetic fields up to 14 T and pulsed magnetic fields up to 60 T at the National High Magnetic Field Laboratory in Los Alamos National Laboratory. For both setups, we used the Akiyama A-probe, typically used for atomic force microscopy \cite{modic-2018-reson-torsion}. This probe consists of a quartz tuning fork which is used to drive an attached Si bridging cantilever into resonance. DC field measurements were obtained with our in-house Quantum Design PPMS Dynacool system in fields up to 14 T. The frequency was tracked using a phase-locked-loop (PLL) via the PLL/PID option available for a Zurich Instruments mid-frequency lock-in amplifier (MFLI), as well as using a custom PID software written in LabVIEW. A driving voltage is applied to one electrical contact on the fork, while the pickup voltage from the vibration is measured on the other. For these measurements, a drive voltage in the range 0.5-5 mV rms was used.

To measure in pulsed field, the cantilever was tuned to the resonant frequency, which varies from 35 kHz to 44 kHz depending upon the cantilever and the mass of the sample. Immediately before the pulse, the drive voltage was shut off and the cantilever-sample system was left to vibrate freely. The $Q$-factors of the cantilevers ranged from 1,000 to 40,000 at base temperature, allowing the cantilever to continue vibrating for the entire duration of the 80 ms field pulse. The resulting waveform was processed using a sliding Fourier transform with a time step ranging from 1-20 \(\mu\)s and window of 500-1000 \(\mu\)s in order to extract the frequency of the sample-cantilever system. Typical drive voltages were on the order of 10 mV rms.

The large magnetic anisotropy of \blio leads to very large frequency shifts for certain crystallographic planes, especially in high magnetic fields, while other planes with little anisotropy may produce almost no shift. This requires focused-ion beam (FIB) milling to create sample sizes that produce the appropriate signal sizes for each case, as shown in \autoref{fig:figure1}.

\subsection{Unit conversion}\label{subsec:conversion}

To convert the measured frequency shift $\Delta f$ to the magnetotropic susceptibility $k$ using Eq.~(\ref{eq:mt-freq}), one would normally require an accurate measurement of the cantilever bending stiffness $S$ at each temperature, 
as well as the sample volume. However, if the magnetic susceptibility is  known  along the relevant crystallographic directions,
the conversion can instead be calibrated using the known magnetic susceptibility anisotropy.

\begin{figure*}[t]
  \centering
  \includegraphics[]{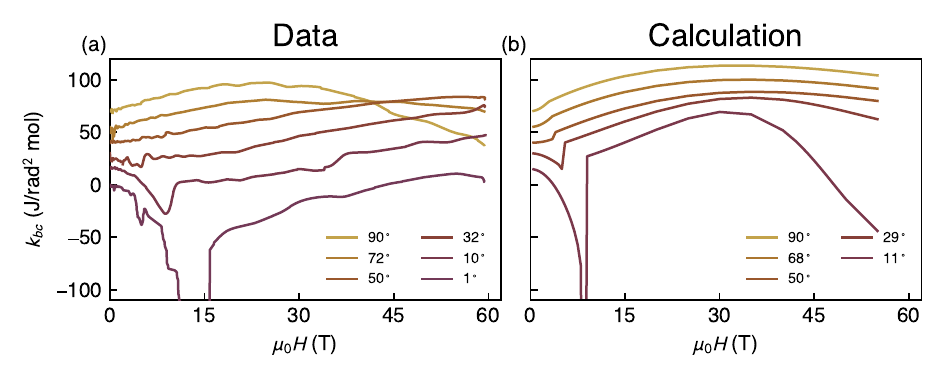}
  \caption{(a) Magnetotropic susceptibility of \blio in the 
  $bc$-plane up to 60~T. The IC transition evolves from 
  $H^\star = 2.8$~T for \hpb to $H^\star \approx 16$~T for \hpc, with $\theta$ defined as the angle between the applied field and the $c$-axis. 
  Above 30~T, the slope of $k$ for \hpb changes sign, suggesting that $g$-factor anisotropy begins to dominate 
  the exchange anisotropy at high magnetic fields. Curves 
  are offset for clarity. (b) Calculated magnetotropic 
  susceptibility for similar field orientations, using the parameters $(J,K,\Gamma)=(0.4,-18,-10)$ meV. The calculated sign change in $k$ for \hpb occurs at a higher field than observed experimentally, indicating that the minimal model overestimates the strength of the zigzag correlations at high fields.}
  \label{fig:bcplane}
\end{figure*}

The magnetic susceptibility of \blio along the $a$, $b$, and $c$ directions is well-characterized \cite{ruiz-2020-high-temper}. As such, we use the low-temperature values of $\chi_a = 0.015$ $\mu_B/$Ir$\cdot$T, $\chi_b = 0.125$ $\mu_B/$Ir$\cdot$T, and $\chi_c  = 0.02$ $\mu_B/$Ir$\cdot$T. In an orthorhombic system, the magnetotropic susceptibility in linear response follows
\begin{equation}
k = (\chi_i-\chi_j)B^2 \cos{2\theta}
\end{equation}
for rotations within the \(ij\)-plane. We fit the measured frequency shift in the low-field limit to 
\begin{equation}
\Delta f = A \cos{2\theta},
\end{equation}
where \(A\) is the amplitude that is proportional to ($\chi_i-\chi_j$). This allows us to find a conversion factor
\begin{equation}
C = \frac{B^2 (\chi_i-\chi_j)}{A}
\end{equation}
to convert the raw frequency shift in Hz to the magnetotropic susceptibility, $k = C \Delta f$, in units of energy per mol without knowledge of the bending stiffness of the cantilever or the sample volume.

 The magnetotropic susceptibility $k=\partial^2F/\partial\theta^2$ has dimensions of energy per angle squared. Treating the radian as a base unit, its natural units are J\,rad$^{-2}$, 
normalized per mole, per unit mass, or per formula unit depending upon context. The magnetic susceptibility reported in Ref.~\cite{ruiz-2020-high-temper} is given in units of $\mu_B$/Ir per tesla, which corresponds to 
magnetotropic susceptibility units of $(\mu_B/\text{Ir})\cdot\text{T}$. To convert to 
molar units, we use
\begin{equation}
k_\text{mol} = k_{\mu_B/\text{Ir}} \cdot 
\frac{N_A \mu_B}{\text{mol}} 
\approx 5.58~\text{J\,mol}^{-1}\text{T}^{-1} 
\times k_{\mu_B/\text{Ir}},
\end{equation}
yielding $k_\text{mol}$ in units of 
J\,mol$^{-1}$\,rad$^{-2}$. Since each formula 
unit of \blio contains one Ir atom, no additional 
factor is needed in the molar conversion.


\section{\label{sec:fields}  Incommensurate order \& high field anisotropy in the $bc$-plane}

\subsection{$H^\star$ across the principal planes}

\autoref{fig:figure2} shows the magnetotropic susceptibility as a function of magnetic field for the three principal crystallographic planes up to 14~T. For $bc$- and $ab$-planes, the low-field response exhibits a transition at $H^\star$, signaling the suppression of the incommensurate (IC) phase. The lowest critical field occurs for \hpb\ at $H^\star=2.8$~T, in quantitative agreement with previous studies~\cite{ruiz-2017-correl-states,li-2020-uncon-magnet}. As the magnetic field is rotated toward the $a$- and $c$-axes (\autoref{fig:figure2}a,b), $H^\star$ shifts rapidly to higher fields, demonstrating the strong anisotropy of the IC phase. Beyond $H^\star$, no additional phase transitions are observed in the $ab$- or $bc$-planes, and the magnetotropic susceptibility evolves smoothly with field. Within the $ac$-plane, (\autoref{fig:figure2}c), no transitions are observed up to 14 T.



 \begin{figure*}[ht!]
  \centering
\includegraphics[]{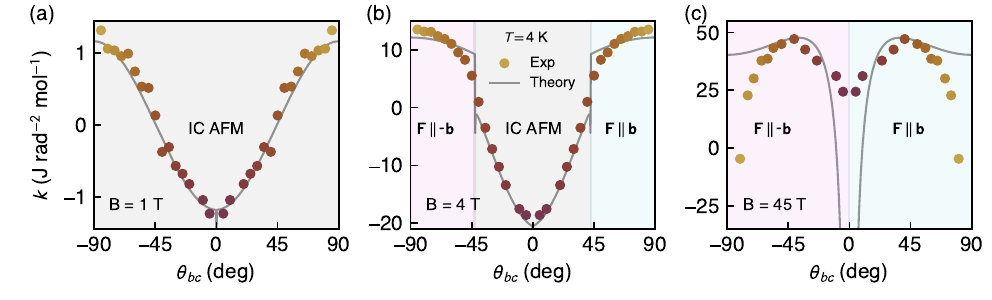}
  \caption{At T = 4 K, angle-dependent magnetotropic susceptibility data of \blio in the $bc$-plane at 1, 4, and 45 T (data symmetrized about \hpc). The
  dark red (\hpc), yellow (\hpb), and intermediate-colored data points correspond to the experimental values of $k$ taken at fixed fields from the data in \cref{fig:figure2} and \cref{fig:bcplane}. The grey lines correspond to the theoretical values from \citet{li-2020-reent-incom}. (a) At 1 T, \blio is in the
IC phase for all field angles (grey shading). The magnetotropic susceptibility versus angle follows the expected $\cos{2\theta}$ behavior.
(b) At 4 T, the IC phase is suppressed for a range of field angles near \hpb (magenta and cyan shading). In this region,
a disordered, yet correlated phase occurs where the ferromagnetic component of spins follows the orientation of the applied magnetic field. (c) At 45 T, the IC order is suppressed for all fields. When the magnetic field approaches the $c$-axis ($\theta = 0^\circ$),
there is a sharp downturn in the magnetotropic susceptibility and it takes on a negative value. The drop in the magnetotropic 
susceptibility near \hpc is evidence for a first-order transition which flips the orientation of the high-field ZZ $\parallel \mathbf{a}$ + FM $\parallel \mathbf{b}$ order parameters, as discussed by \citet{li-2020-reent-incom}. At higher fields, the drop in $k$ for fields
near the $c$-axis diminishes and instead, $k$ becomes negative near $\theta = \pm 90^\circ$, consistent with a change in the sign of $\chi_b - \chi_c$ at high fields.}
  \label{fig:figure5}
\end{figure*}

\subsection{\label{sec:highfieldbc}Anisotropy in the $bc$-plane up to 60 tesla}

\autoref{fig:bcplane} shows the measured (a) and calculated (b) magnetotropic susceptibility for several field orientations in the $bc$-plane up to 60~T. At low magnetic fields, the magnetotropic susceptibility is quadratic in field, with a coefficient proportional to the magnetic susceptibility anisotropy $(\chi_b-\chi_c)$ in the plane of rotation. The amplitude follows the expected $\cos2\theta$ dependence, with the maximum (minimum) corresponding to the magnetically easy (hard) axis (\autoref{fig:figure5}a). As the field is rotated toward \hpc, the critical field increases to $H^\star\approx16$~T.

Previous theoretical work showed that, for magnetic field applied along the $c$-axis, the state above $H^\star$ in the $bc$-plane is characterized by a zigzag order parameter along the $a$-axis and a ferromagnetic order parameter along the $b$-axis~\cite{li-2020-uncon-magnet} (\autoref{subsec:HighFieldConf}).
For \hpc, the intermediate-field phase arises from the spontaneous selection of one of the two possible order parameter orientations, with transition into the field-polarized phase $H^{\star\star}$ predicted at $\sim45$~T~\cite{li-2020-reent-incom}.

In \autoref{fig:bcplane}a, no discernible signature of $H^{\star\star}$ is observed in the $bc$-plane. This is consistent with our symmetry analysis: the transition is expected only for \hpc, where the discontinuity in the magnetotropic susceptibility vanishes because the jump in $k$ is proportional to the square of the slope of the $H-\theta$ phase boundary, which is zero along this high-symmetry direction. Additionally, it would only be possible to observe this transition for field precisely aligned with the $c$-axis. For all other field orientations, even those infinitesimally away from the $c$-axis, the magnetic field breaks the symmetry and uniquely selects the ferromagnetic/zigzag domain, and the evolution toward the field-polarized state therefore proceeds as a crossover. 

The high-field magnetotropic response in the $bc$-plane appears to develop according to how the magnetic susceptibility tensor evolves with temperature, reflecting a competition between exchange-driven and $g$-factor-driven anisotropy. 
Above $\sim35$~T, the slope of the magnetotropic susceptibility, which reflects the sign of $\chi_b-\chi_c$, changes sign for \hpb. 
At high temperature, where the $g$-tensor governs the anisotropy, the principal susceptibilities order as $\chi_c > \chi_b > \chi_a$~\cite{ruiz-2020-high-temper}; upon cooling, exchange anisotropy reverses this ordering in the $bc$-plane below 100~K, with $\chi_b > \chi_c$ at low temperature~\cite{Takayama2015,ruiz-2020-high-temper}. The sign change observed above 40~T for \hpb is consistent with the Zeeman energy reasserting the high-temperature ordering, suggestive of an intermediate field regime---above $H^\star$ but below saturation---where $g$-factor anisotropy \textit{and} exchange anisotropy govern the magnetic response.


\subsection{Angle-dependent evidence for the high-field correlated regime}

Evidence for the high-field correlated regime above $H^\star$ can be seen more clearly in the angle-dependent data in \autoref{fig:figure5}. At 1~T (panel a), \blio is in the IC phase for all field angles and the magnetotropic susceptibility follows the $\cos 2\theta$ dependence expected of the linear response regime ($M_i=\chi_{ij}H_j$). At 4~T (panel b), the IC phase is suppressed for fields near the $b$-axis, and entrance into the AFM phase occurs at $\theta\sim\pm45^\circ$ as the field rotates towards the $c$-axis.

At 45~T (panel c), the IC phase is suppressed for all field orientations. When the magnetic field approaches the $c$-axis, there is a sharp downturn in $k$, consistent with a high-field reversal of the magnetic torque $k=\partial\tau/\partial\theta$. For $H>H^\star$ on either side of the $c$-axis, the favored spin configuration is fixed by the Zeeman coupling to the ferromagnetic component induced by the field projection along the $b$-axis. As the field angle crosses the $c$-axis, the sign of this projection changes, forcing a discontinuous reorientation of the ferromagnetic component and producing a first-order angular transition. Consistent with this expectation, the torque abruptly switches sign at \hpc. Similar behavior was observed in \glio at high magnetic fields~\cite{modic-2014-realiz-three,modic-2017-robus-spin,riedl-2019-sawtoot-torque} and was predicted in \blio~\cite{li-2020-reent-incom}, but has not been observed experimentally until now. The strong drop near the $c$-axis indicates that the spins remain strongly correlated up to at least 45~T, and that higher magnetic fields are needed to enter the field-polarized phase.


At 45 T, $k$ becomes negative near $\theta = \pm 90^\circ$ (\hpb), consistent with a change in sign of $\chi_b - \chi_c$ at high fields. This could reflect $g$-tensor anisotropy, a field-driven reorganization of the exchange anisotropy, or a combination of both. Notably, the minimal $J$-$K$-$\Gamma$ model predicts this sign change at higher fields than observed, overestimating the stability of the exchange-driven anisotropy at high fields.


\section{Theoretical Framework}\label{sec:theory}
  
The experimental results discussed in this work reveal a pronounced anisotropy of the field-induced phases, including the emergence of an additional high-field transition in the $ac$-plane that is absent in the $ab$- and $bc$-planes. This section outlines the theoretical framework used throughout the paper, including all calculations underlying the theoretical curves compared with experiment. We first introduce the minimal $J$-$K$-$\Gamma$ model, then derive the effective high-field description and symmetry constraints that determine the allowed ordered states, and finally analyze the instability leading to the transition at $H^{\star\star}$.

\subsection{Minimal $J$-$K$-$\Gamma$ model}
\label{subsec:minimalmodel}

The magnetic properties of \blio are described by 
spin-orbit entangled $j_\text{eff}=1/2$ moments on a hyperhoneycomb lattice with dominant bond-dependent exchange interactions~\cite{ducatman-2018-magnet-struc,rousochatzakis-2018-magnet-field,lee-2015-theor-magnet,lee-2016-two-iridat}. We adopt the minimal nearest-neighbor $J$-$K$-$\Gamma$ Hamiltonian with an anisotropic Zeeman coupling:
\begin{equation}
    \mc{H}=\sum_t\sum_{\langle ij\rangle\in t}
    \mc{H}_{ij}^t+\mc{H}_{Z},
\end{equation}
where $\mc{H}_{ij}^t$ is the nearest-neighbor 
$J$-$K$-$\Gamma$ Hamiltonian
\begin{equation}\label{eq:jkg}
    \mc{H}_{ij}^t=J\vec{S}_i\cdot\vec{S}_j
    +KS_i^{\alpha_t}S_j^{\alpha_t}
    +\sigma_t\Gamma(S_i^{\beta_t}S_j^{\gamma_t}
    +S_i^{\gamma_t}S_j^{\beta_t}),
\end{equation}
and $\mc{H}_Z$ is the Zeeman term
\begin{equation}\label{eq:HZ1}
    \mc{H}_{Z}=-\mu_B\vec{H}\cdot\sum_{i}\vec{g}_i
    \cdot\vec{S}_i.
\end{equation}
In the exchange part of the Hamiltonian, $t$ labels symmetry-distinct Ir-Ir bonds and $\alpha_t$, $\beta_t$, and $\gamma_t$ denote bond-dependent spin components. The site-dependent $g$-tensor contains uniform diagonal components and an alternating off-diagonal element $g_{ab}$ that changes sign between the two zigzag chain families. Parameters consistent with prior studies are used for comparison with our experiments~\cite{ruiz-2017-correl-states, li-2020-uncon-magnet}.

\subsection{Magnetic configuration for $H\!\geq\!H^\star$}
\label{subsec:HighFieldConf}
The zero-field ground state is an incommensurate 
multi-sublattice structure. When the magnetic field exceeds $H^\star$, the incommensurate modulation is suppressed and the spin configuration simplifies. All spins belonging to one set of zigzag chains align along a common direction $\bf{A}$, while spins on the complementary chains align along $\bf{A'}$ { (Figure~1 in  the main  text)}.
The magnetic state can therefore be expressed in terms of two collective vectors,
\begin{align}
\mathbf{F}=\frac{\mathbf{A'}+\mathbf{A}}{2},\qquad
\mathbf{G}=\frac{\mathbf{A'}-\mathbf{A}}{2}.
\end{align}
Here, $\bf{F}$ represents the uniform (net) magnetization and $\bf{G}$ represents a residual staggered (also termed `zigzag') component. The spin-length constraints impose
\begin{align}\label{eq:constraints}
\mathbf{F}^2+\mathbf{G}^2=S^2,\qquad
\mathbf{F}\cdot\mathbf{G}=0.
\end{align}
The high-field magnetic state is thus fully characterized by the symmetry-allowed components of $\bf{F}$ and $\bf{G}$. 

\begin{table*}[t!]
\centering
\renewcommand{\arraystretch}{1.35}
\begin{tabular*}{0.85\textwidth}{@{\extracolsep{\fill}} 
c c c c c c c}
\toprule
 & $C_{2\mathbf{a}}$ & ${\Theta}C_{2\mathbf{a}}$ 
 & $C_{2\mathbf{b}}$ & ${\Theta}C_{2\mathbf{b}}$ 
 & $C_{2\mathbf{c}}$ & ${\Theta}C_{2\mathbf{c}}$ \\
\toprule
$F_a$ & $+$ & $-$ & $-$ & $+$ & $-$ & $+$ \\
$F_b$ & $-$ & $+$ & $+$ & $-$ & $-$ & $+$ \\
$F_c$ & $-$ & $+$ & $-$ & $+$ & $+$ & $-$ \\
\midrule
$G_a$ & $-$ & $+$ & $+$ & $-$ & $-$ & $+$ \\
$G_b$ & $+$ & $-$ & $-$ & $+$ & $-$ & $+$ \\
$G_c$ & $+$ & $-$ & $+$ & $-$ & $+$ & $-$ \\
\toprule
$\mathbf{H}\parallel a$ & 
{$\checkmark(F_b,F_c,G_a)$} &{}&{} &
{$\checkmark(F_b,G_a,G_c)$} &{}&
{$\checkmark(F_c,G_c)$} \\
$\mathbf{H}\parallel b$ &{} &
{$\checkmark(F_a,G_b,G_c)$} &
{$\checkmark(F_a,F_c,G_b)$} &{}&{}&
{$\checkmark(F_c,G_c)$} \\
$\mathbf{H}\parallel c$ &{} &
{$\checkmark(F_a,G_b,G_c)$} &{} &
{$\checkmark(\textbf{\textit{F}}_\textbf{\textit{b}},\textbf{\textit{G}}_\textbf{\textit{a}},G_c)$} &
{$\checkmark(F_a, \textbf{\textit{F}}_\textbf{\textit{b}},\textbf{\textit{G}}_\textbf{\textit{a}},G_b)$} &{} \\
\midrule
$\mathbf{H}\parallel ab$ &{}&{}&{}&{}&{}&
{$\checkmark(F_c,G_c)$} \\
$\mathbf{H}\parallel bc$ &{} &
{$\checkmark(F_a,G_b,G_c)$} &{}&{}&{}&{}\\
$\mathbf{H}\parallel ac$ &{}&{}&{}&
{$\checkmark(\textbf{\textit{F}}_{\textbf{\textit{b}}}, \textbf{\textit{G}}_{\textbf{\textit{a}}}, \textbf{\textit{G}}_{\textbf{\textit{c}}})$}&{}&{}\\
\bottomrule
\end{tabular*}
\caption{
Rows 2-7: Transformation properties of $F_\nu$ and $G_\nu$ ($\nu=a,b,c$) under the symmetry operations listed in row 1. A $+$ ($-$) indicates that the component is even (odd) under the operation. Note that, unlike $C_{2\mathbf{c}}$, the operations $C_{2\mathbf{a}}$ and $C_{2\mathbf{b}}$ exchange the two zigzag chain families. 
Rows 8-13: Check marks indicate the symmetries of the Hamiltonian when the field is applied along a high-symmetry axis (rows 8-10) or a principal plane (rows 11-13). The components listed in the parentheses next to the tick marks are odd under these symmetries---these are the only components capable of ordering spontaneously. Components that order spontaneously above $H^\star$ (as determined by classical energy minimization) are bolded.}\label{tab:full_symmetry}
\end{table*}

\subsection{Energy in terms of ${\bf F}$ and ${\bf G}$}\label{subsec:freenergy}

To make the couplings responsible for the field-induced instabilities more transparent, we express the energy for $H\!\geq\!H^\star$ in terms of $\mathbf{F}$ and $\mathbf{G}$:
\be\label{eq:EvsFG}
\!\!\!\mc{H}\!=\!\frac{1}{2}\!\!\sum_{\nu=a,b,c}\!\!\!
\left(f_\nu F_\nu^2\!+\!g_\nu G_\nu^2\right)
\!-\!\sqrt{2}\Gamma(F_aG_b\!+\!F_bG_a)
\!+\!\mc{H}_{Z},
\ee
where the coefficients $f_\nu$ and $g_\nu$ depend on the microscopic exchange parameters according to
\be
\ra{1.25}
\begin{array}{ll}
f_a=K+3J-\Gamma\,, & g_a=K+J+\Gamma\,,\\
f_b=K+3J\,, & g_b=-K+J\,,\\
f_c=K+3J+\Gamma\,, & g_c=K+J-\Gamma\,.
\end{array}
\ee
The Zeeman energy takes the form
\begin{align}\label{eq:HZ2}
\!\!\mc{H}_{Z} = -\mu_B\Big[&
H_a(g_{aa}F_a\!-\!g_{ab}G_b)
\!+\!H_b(g_{bb}F_b\!-\!g_{ab}G_a)\notag\\
&+ H_c g_{cc}F_c\Big]\,.
\end{align}

The crucial feature of  Eq.~(\ref{eq:EvsFG}) is the $\Gamma$-dependent cross-coupling term $-\sqrt{2}\Gamma(F_aG_b+F_bG_a)$, which mixes uniform and zigzag components. There are two main consequences of such cross-coupling terms: i) a field-induced magnetization (e.g., $F_b$ for ${\bf H}\parallel {\bf b}$) can drive a significant zigzag component in the transverse direction ($G_a$)~\cite{rousochatzakis-2018-magnet-field}; ii) for components of ${\bf F}$ and ${\bf G}$ that transform non-trivially under a symmetry of the Hamiltonian, e.g., $F_b$ and $G_a$ for fields in the $ac$-plane, a cross-coupling term $\propto F_b G_a$ can give rise to an instability, which is exactly what we report in this study. The following subsection provides a more systematic analysis of these points based on symmetry.

\subsection{Symmetry considerations}
\label{subsec:crystalsymmetry}

Apart from primitive translations, the zero-field 
Hamiltonian of \blio is invariant under six operations: $C_{2a}$, ${\Theta}C_{2a}$, $C_{2b}$, ${\Theta}C_{2b}$, $C_{2c}$, and ${\Theta}C_{2c}$, where $C_{2\nu}$ denotes a two-fold rotation around axis $\nu$ and ${\Theta}$ denotes time-reversal. An applied magnetic field explicitly breaks a subset of these symmetries depending on its orientation. The remaining unbroken symmetries constrain which components of $\bf{F}$ and $\bf{G}$ can appear spontaneously, and therefore determine the structure of the allowed intermediate-field phases, if any. The full transformation properties and allowed order parameters for each field orientation are summarized in \autoref{tab:full_symmetry}.

To read \autoref{tab:full_symmetry}: the $+/-$ signs in rows 2-7 indicate whether each order parameter component is even ($+$) or odd ($-$) under the corresponding symmetry operation. For a given field orientation, only the operations that remain unbroken are relevant (indicated by check marks in rows 8-13). A component can order spontaneously only if it is odd ($-$) under any subset of the symmetry group of the Hamiltonian.  
So, according to the last three rows of \autoref{tab:full_symmetry}, the only possible components that can, {\it in principle}, appear spontaneously are: i) $F_c$ and $G_c$ for fields in the $ab$-plane, ii) $F_a$, $G_b$ and $G_c$  for fields in the $bc$-plane, and iii) $F_b$, $G_a$ and $G_c$ for fields in the $ac$ plane.

\begin{figure*}[t!]
\centering
\includegraphics[width=0.99\textwidth,trim=0cm 0cm 0cm 0cm, clip=true]{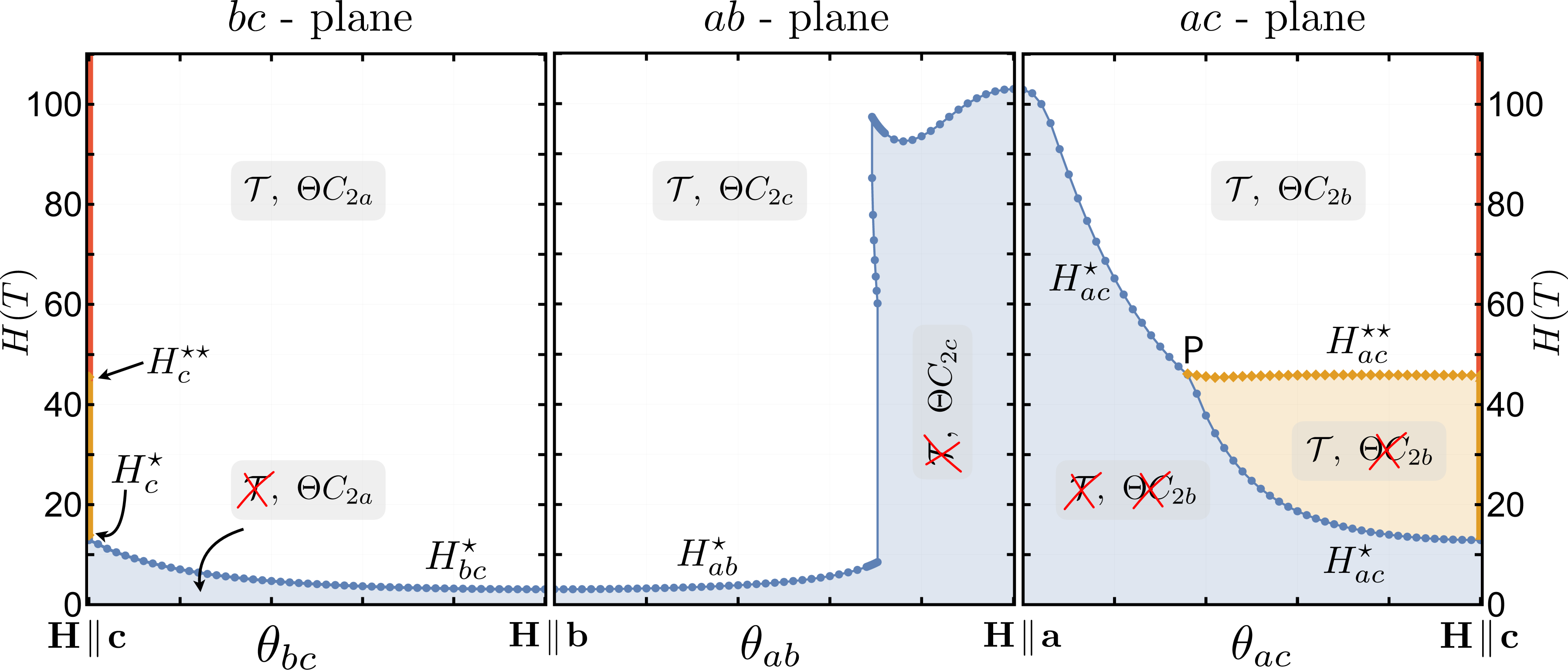}
\caption{Phase diagram for fields in the $bc$-, $ab$- and $ac$-planes as obtained from numerical classical energy minimizations, with parameters $(J,K,\Gamma)=(-0.4,-18,-10)$\,meV, $g_{aa}\!=\!g_{bb}\!=\!g_{cc}\!=\!2$ and $g_{ab}\!=\!0$. The symmetries that are broken in each phase are indicated by a cross mark, $\mc{T}$ denotes the translation group, and ${\Theta}$ the time-reversal operation. The intermediate Ising-like phase found experimentally in the present work for fields in the $ac$ plane (rightmost panel) is absent from the phase diagram published in Ref.~\cite{li-2020-reent-incom}. The red line above $H_c^{\star\star}$ indicates the {\it fully} polarized phase.}
\label{fig:PD1}
\end{figure*}

\section{Theoretical results}

\subsection{Key aspects of the phase diagram}

Figure~\ref{fig:PD1} shows the theoretical phase diagram of $\beta$-Li$_2$IrO$_3$ for fields in the $bc$-, $ab$- and $ac$-planes as obtained from numerical classical energy minimizations, based on the $J$-$K$-$\Gamma$ model and the general six-sublattice ansatz established in previous works~\cite{ducatman-2018-magnet-struc,rousochatzakis-2018-magnet-field,li-2020-uncon-magnet,li-2020-reent-incom}. 
As discussed in the main text, the main difference compared to the phase diagram reported previously by ~\citet{li-2020-reent-incom} is the intermediate phase in the $ac$-plane. 
The main aspects of the phase diagram can be summarized as follows:

\begin{itemize}

\item For fields in the $bc$-plane, the IC phase does not break ${\Theta}C_{2a}$, and therefore $F_a$, $G_b$ and $G_c$ vanish in the IC phase.

\item For fields in the $ab$-plane, the IC phase does not break ${\Theta}C_{2c}$, and therefore $F_c$ and $G_c$ vanish in the IC phase.

\item For fields in the $ac$-plane, ${\Theta}C_{2b}$ is broken in both the IC and the intermediate phase. Hence, $F_b$, $G_a$ and $G_c$ are nonzero in both phases.

\item For fields along the $a$-axis, the Hamiltonian is invariant under ${\Theta}C_{2b}$, ${\Theta}C_{2c}$ and $C_{2a}$. The IC phase breaks ${\Theta}C_{2b}$ and $C_{2a}$ but not ${\Theta}C_{2c}$~\cite{li-2020-uncon-magnet}. Therefore, in the IC phase, $F_c$ and $G_c$ vanish, whereas $F_b$ and $G_a$ are nonzero.

\item For fields along the $b$-axis, the Hamiltonian is invariant under ${\Theta}C_{2a}$, ${\Theta}C_{2c}$ and $C_{2b}$. The IC phase does not break any of these symmetries~\cite{li-2020-uncon-magnet}, so $F_a$, $G_b$, $G_c$ and $F_c$ all vanish in the IC phase.

\item For fields along the $c$-axis, the Hamiltonian is invariant under ${\Theta}C_{2a}$, ${\Theta}C_{2b}$ and $C_{2c}$. The IC and the intermediate phase break ${\Theta}C_{2b}$ and $C_{2c}$ but not ${\Theta}C_{2a}$~\cite{li-2020-uncon-magnet}. Hence, in both of these phases, $F_a$, $G_b$ and $G_c$ vanish, whereas $F_b$ and $G_a$ are nonzero.

\end{itemize}

Furthermore, the nature of the various transitions are as follows:

\begin{itemize}

\item For fields in the $bc$ plane, the IC-PM transition is continuous everywhere. 

\item For fields in the $ab$ plane, the IC-PM transition is continuous only for $\theta_{ba}\leq 0.35\pi$. 

\item For fields in the $ac$ plane, both the IC-PM and IC-intermediate phase ($F_b + G_a$ order) transitions are first-order, whereas the Ising-PM transition is continuous. 

\item For fields in the $ac$ plane, the Ising-PM line terminates, on one side, at the critical end point `P', when it meets the $H_{ac}^\star$ first-order lines. On the other side, the Ising-PM boundary terminates at $H_c^{\star\star}$ at ${\bf H}\!\parallel\!{\bf c}$. 

\end{itemize}

\subsection{Dependence of the phase boundaries on exchange parameters}

\begin{figure*}[t!]
  \centering
\includegraphics[width=0.99\textwidth,trim=0cm 0cm 0cm 0cm, clip=true]{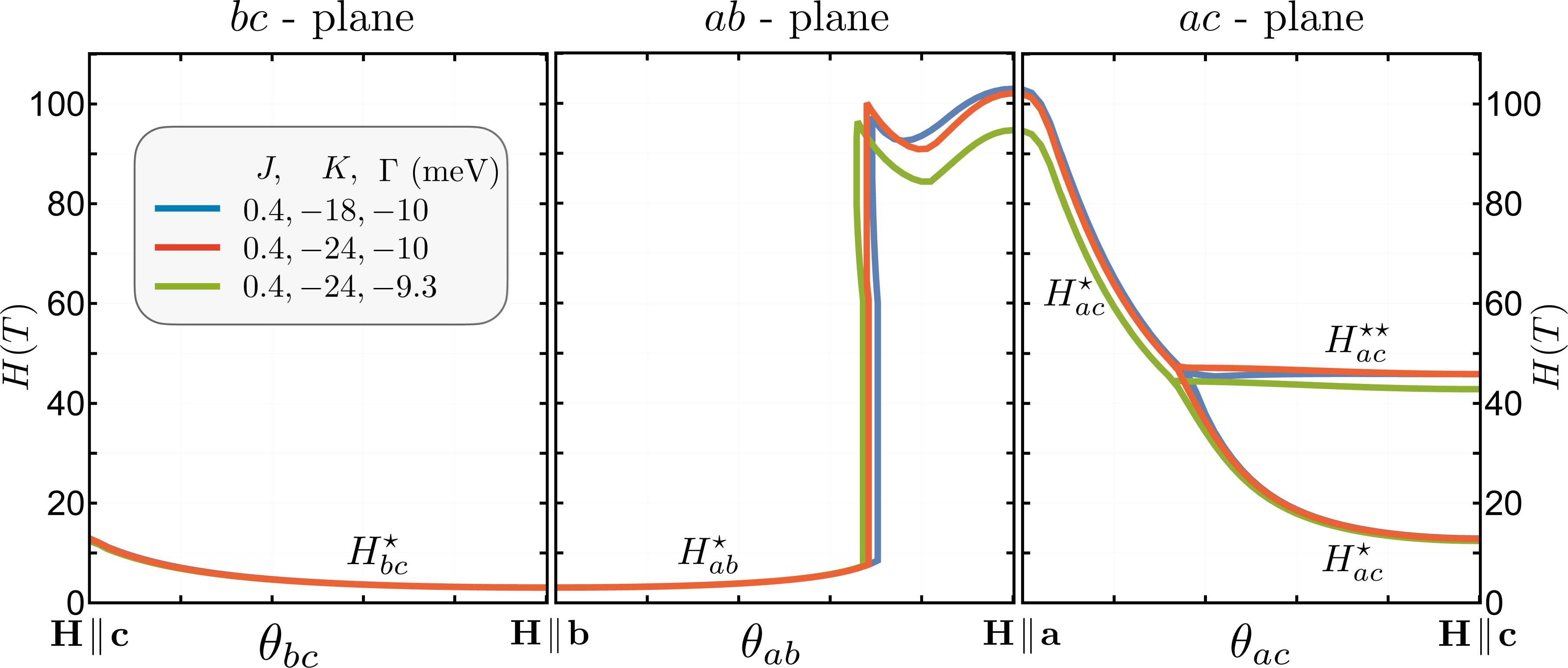}
\caption{Theoretical phase diagram of $\beta$-Li$_2$IrO$_3$ obtained by classical energy minimizations within the $J$-$K$-$\Gamma$ model for three different sets of exchange parameters (see inset), and for an isotropic $g$-tensor with $g=2$.}
\label{fig:PDvspars}
\end{figure*}

Figure~\ref{fig:PDvspars} shows the theoretical phase diagram for three different sets of exchange parameters $(J, K, \Gamma)$: i) $(0.4, -18, -10)$~meV, which is the parameter set used in Ref.~\cite{li-2020-reent-incom}; ii) $(0.4, -24, -10)$~meV, which differs from the first set only in the value of the dominant coupling $K$; and iii) $(0.4, -24, -9.3)$~meV, which is the refined parameter set obtained in  Ref.~\cite{Halloran2022}. The comparison between sets i) and ii) shows that the phase boundaries show very little dependence on the Kitaev coupling. By contrast, a comparison between (i)-(ii) and (iii) shows that reducing the value of $\Gamma$ lowers $H_{ab}^\star$ for $\theta_{ab}\gtrsim0.35\pi$, $H_{ac}^\star$ for $\theta_{ac}\lesssim0.2\pi$, and $H_{ac}^{\star\star}$. Finally, $H_{bc}^\star$ and $H_{ab}^\star$ for $\theta_{ab}\lesssim0.35\pi$ are unaffected by $K$ and $\Gamma$, so they only depend on $J$. These results are consistent with the expressions (\ref{eq:hstar}) and (\ref{eq:hcdstar}) discussed below.

\subsection{Critical field formulas}
\label{subsec:criticalfields}

The theoretical values of the critical fields, calculated directly from the exchange parameters~\cite{li-2020-uncon-magnet}, are:
\begin{align}\label{eq:hstar}
\mu_B H_a^\star &\simeq(0.54J+0.57|\Gamma|)
    \frac{4S}{g_{aa}},\nonumber\\
\mu_B H_b^\star &\simeq 0.42J\left(\frac{4S}{g_{bb}}\right),\\
\mu_B H_c^\star &\simeq(0.94J+0.04|\Gamma|)
    \frac{4S}{g_{cc}}.\nonumber
\end{align}
Here $S=1/2$ and $(g_{aa},g_{bb},g_{cc})$ are the diagonal elements of the $\mathbf{g}$-tensor. Explicit $g$-factor values are not available in the literature for \blio. Following \citet{majumder-2019-anisot-temper}, 
which reports effective moments from Curie-Weiss fits ($\mu_a=1.64$, $\mu_b=1.65$, $\mu_c=1.74$), we estimate
\begin{equation}
g_{cc}=\frac{1.74}{1.65}\times2\approx2.1,\qquad
g_{aa}=\frac{1.64}{1.65}\times2\approx2.0,
\end{equation}
taking $g_{bb}=2$ as reference, consistent with the quantitative agreement obtained for $H_b^\star$. Including this moderate $g$-tensor anisotropy slightly reduces the predicted values of $H_c^\star$ and $H_c^{\star\star}$ but does not fully account for the remaining quantitative discrepancy, suggesting either a small residual field misalignment or limitations of the minimal model at this level of precision.

Additionally, for \hpc the high-field transition 
$H^{\star\star}$ 
depends on both $J$ and $\Gamma$:
\begin{equation}\label{eq:hcdstar}
\mu_B H_c^{\star\star}\simeq\left(\Gamma+2J
+\sqrt{(\Gamma-2J)^2+8\Gamma^2}\right)
\frac{S}{2g_{cc}}.
\end{equation}
Importantly, the critical fields $H^\star$ and 
$H^{\star\star}$ depend on $J$ and $\Gamma$ but are independent of the Kitaev coupling $K$ at leading order. This is because, within the semi-analytical description, $K$ primarily sets the overall energy scale of the background IC state but drops out of the instability conditions that determine the critical fields. Concretely, $H_b^\star$ is independent of both $K$ and $\Gamma$, while $H_a^\star$, $H_c^\star$, and $H_c^{\star\star}$ depend on $J$ and $\Gamma$ and are essentially insensitive to $K$. Physically, this means that the anisotropic field-induced collapse of IC order and the subsequent high-field transitions are governed by the subleading interactions $J$ and $\Gamma$ that select and mix the uniform ferromagnetic and zigzag components, not by the dominant Kitaev interaction itself.

\subsection{Stability analysis (which planes host an intermediate phase, and why)}\label{subsec:LandauAnalysis1}

Here we address the question on why there is an intermediate phase for fields in the $ac$-plane (for a certain range of field orientations), but not for fields in the $bc$- and $ab$-planes. While the absence of an intermediate phase in the $ab$-plane can be attributed to the absence of a cross-coupling term $\propto{F_c G_c}$, the situation in the $bc$-plane is not immediately evident, because the energy contains a cross-coupling term $-\sqrt{2}\Gamma F_a G_b$ which, at first sight,  can favor nonzero $(F_a, G_b, G_c)$ and break the mirror symmetry ${\Theta}C_{2a}$. As discussed in the main text, the absence of such order is related to the fact that the mirror symmetry is intact in the low-field IC phase, and therefore an intermediate phase would imply two (unrelated) instabilities happening at the same field $H_{bc}^\star$, which is not generic (see main text). A more direct analysis of  why there is  spontaneous symmetry breaking above $H^\star$ in the $ac$-plane but not in the $bc$-plane is to take a closer look at  the corresponding instability modes.

To keep things simple, we take $g_{ab}\!=\!0$ and $g_{aa}\!=\!g_{bb}\!=\!g_{cc}$. We also denote $h_\nu\!\equiv\!g \mu_B H_\nu$, $\nu=a, b, c$.
Let us use a convenient parametrization of ${\bf F}$ and ${\bf G}$ that satisfies automatically the constraints (\ref{eq:constraints}). We introduce the `spherical' unit vectors (expressed in the $abc$-frame)
\be
\ra{1.25}
\begin{array}{l}
{\bf e}_r = (\sin{\theta} \cos{\phi}, \sin{\theta} \sin{\phi}, \cos{\theta})\,,\\
{\bf e}_\theta = (\cos{\theta} \cos{\phi}, \cos{\theta}\sin{\phi}, -\sin{\theta})\,,\\
{\bf e}_\phi = (-\sin{\phi}, \cos{\phi}, 0)\,,
\end{array}
\ee
and parametrize 
\be\label{eq:paramaterization}
{\bf F}\!=\!S\cos{r}~{\bf e}_r\,,~~
{\bf G}\!=\!S\sin{r}(\sin{\alpha} ~{\bf e}_\theta\!+\!\cos{\alpha}~{\bf e}_\phi).
\ee

We now study the stability of the high-field paramagnetic (PM) phase as the magnetic field is reduced for a fixed field direction. The onset of a phase transition is determined by the stability of the stationary solution of the energy functional. To this end, we expand the energy to second order in small fluctuations around the stationary point.
Let 
\be
\overline{\bf X}=(\overline{\alpha},\overline{\phi},\overline{\theta},\overline{r})
\ee
denote the stationary solution at a given magnetic field,  and consider the fluctuations
\begin{align}
\delta\mathbf{X}
=
(\delta\alpha,\delta\phi,\delta\theta,\delta r),
\end{align}
where $\delta\alpha=\alpha-\overline{\alpha}$, $\delta\phi=\phi-\overline{\phi}$, $\delta\theta=\theta-\overline{\theta}$ and 
$\delta r=r-\overline{r}$.  For fields in the $ac$- and $bc$-planes, we find that the corresponding change in the energy, $\delta E = E-\overline{E}$,  up to quadratic order in the fluctuations, takes the `block-diagonal' form
\begin{align}
\frac{\delta E}{S^2}
=
\frac12
\delta\mathbf{X}
\cdot
\begin{pmatrix}
{\bf A} & 0\\
0 & {\bf B}
\end{pmatrix}
\cdot
\delta\mathbf{X}^{T}\,.
\label{eq:quadratic}
\end{align}
Hence, the sector $(\delta\alpha, \delta\phi)$ decouples from $(\delta\theta, \delta r)$. The instability is governed by the $(\delta\alpha, \delta\phi)$ sector. In particular, the critical field (if any) corresponds to the vanishing of the lowest eigenvalue of ${\bf A}$. 
The explicit forms of the $2\times2$ matrices ${\bf A}$ and ${\bf B}$ depend on the magnetic-field orientation (see below). \\

\noindent{\bf $ac$-plane:}\\
For fields in the $ac$-plane and $H\geq H^{\star\star}_{ac}$, the stationary point equations are characterized by
\be\label{eq:StatCond2}
\ra{1.25}
\begin{array}{ccc}
\overline{F}_a \neq 0\,,& \overline{F}_c\neq0\,,
&\overline{G}_b\neq0\,,
\\
\overline{F}_b =
0\,,&\overline{G}_a=0\,,&\overline{G}_c=0\,,
\end{array}
\ee
where $\overline{F}_a$ is the value of $F_a$ at the stationary point, and similarly for the other components. In terms of parametrization (\ref{eq:paramaterization}), the PM stationary point can be described by
\be
\overline{\phi}=0\,,~~\overline{\alpha}=0\,,
\ee
whereas $\overline{\theta}({\bf H})$ and $\overline{r}({\bf H})$ are determined by the stationary conditions and depend on the magnitude and the direction of ${\bf H}$.  
The elements of the matrices ${\bf A}$ and ${\bf B}$ of Eq.~(\ref{eq:quadratic}) are
\be\label{eq:HessianA}
\!\!\!
\ra{1.25}
\begin{array}{l}
A_{\alpha\alpha} \!=\! 
\overline{G}_b^2 [
(g_c\!-\!g_b) \overline{F}_a^2
\!+\!(g_a\!-\!g_b)\overline{F}_c^2
]/\overline{{\bf F}}^2\!+\!\sqrt{2}\Gamma \overline{F}_a \overline{G}_b\,, 
\\
A_{\phi\phi} \!=\! 
(f_b\!-\!f_a)\overline{F}_a^2
\!+\!(g_a\!-\!g_b)\overline{G}_b^2
\!+\!h_a\overline{F}_a
\!+\!4\sqrt{2}\Gamma \overline{F}_a \overline{G}_b
\,,
\\
A_{\alpha\phi}\!=\!
[(g_b\!-\!g_a)\overline{G}^2_b\!-\!2\sqrt{2}\Gamma \overline{F}_a\overline{G}_b]\cos\overline{\theta}\,,
\end{array}
\ee
and
\be
\!\!\!
\ra{1.25}
\begin{array}{l}
B_{\theta\theta}\!=\!(f_a\!-\!f_c)(\overline{F}_c^2\!-\!\overline{F}_a^2)\!+\! {\bf h}\cdot\overline{\bf F}\!+\! \sqrt{2}\Gamma \overline{F}_a\overline{G}_b\,,
\\
B_{rr}\!=\!
[
(g_b\!-\!f_a)\overline{F}_a^2
\!+\!
(g_b\!-\!f_c)\overline{F}_c^2
][1-\frac{\overline{G}_b^2}{\overline{\bf F}^2}]
\!+\!
{\bf h}\cdot\overline{\bf F}\\
~~~~~~~~+ 
4\sqrt{2}\Gamma \overline{F}_a\overline{G}_b
\,,
\\
B_{{\theta}r} \!=\!
[2(f_c\!-\!f_a) \overline{F}_a \overline{G}_b \!-\!\sqrt{2} \Gamma (\overline{\bf F}^2\!-\!\overline{G}_b^2) ]\cos\overline{\theta}
\\
~~~~~~~~
+\overline{G}_b (h_a \cos\overline{\theta}
\!-\!h_c \sin\overline{\theta})
\,,
\end{array}
\ee
where $\overline{\bf F}=(\overline{F}_a, 0, \overline{F}_c)$ and ${\bf h}=(h_a, 0, h_c)$. The instability takes place when the lowest eigenvalue of ${\bf A}$ 
vanishes. This condition gives the value of $H_{ac}^{\star\star}$. If this is above $H_{ac}^\star$ then there is an intermediate phase. If, on the other hand, this value coincides with  $H_{ac}^\star$ then we enter the IC phase directly (where, in addition to the uniform order parameters $F_b$, $G_a$ and $G_c$, we have the modulated order parameters corresponding to translation symmetry breaking), and no intermediate phase exists.

\begin{figure}[t!]
\centering
\includegraphics[width=0.99\columnwidth,trim=0cm 0cm 0cm 0cm, clip=true]{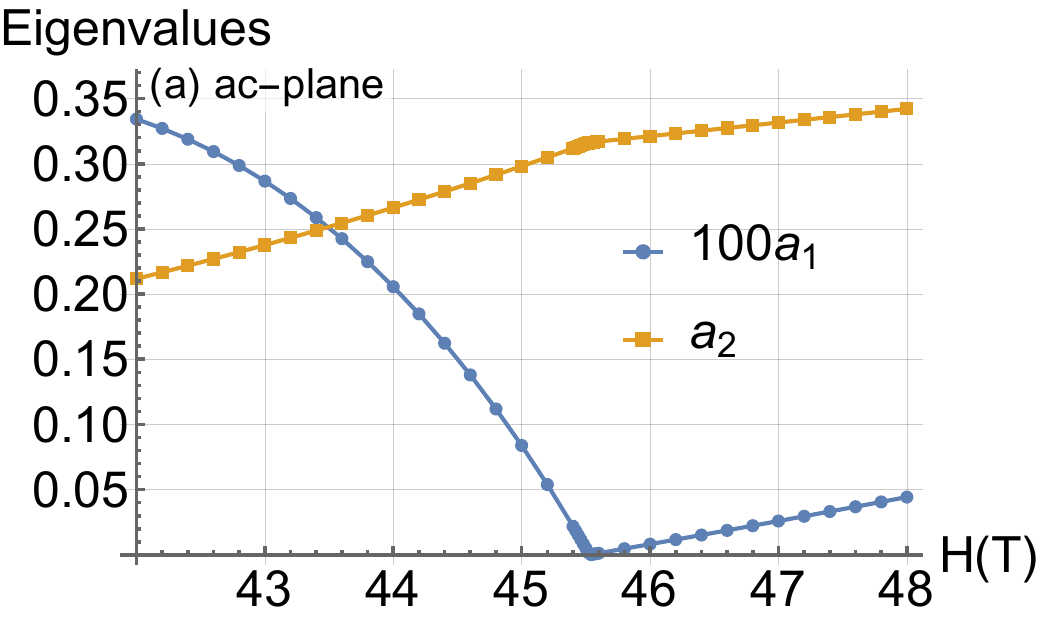}
\includegraphics[width=0.99\columnwidth,trim=0cm 0cm 0cm 0cm, clip=true]{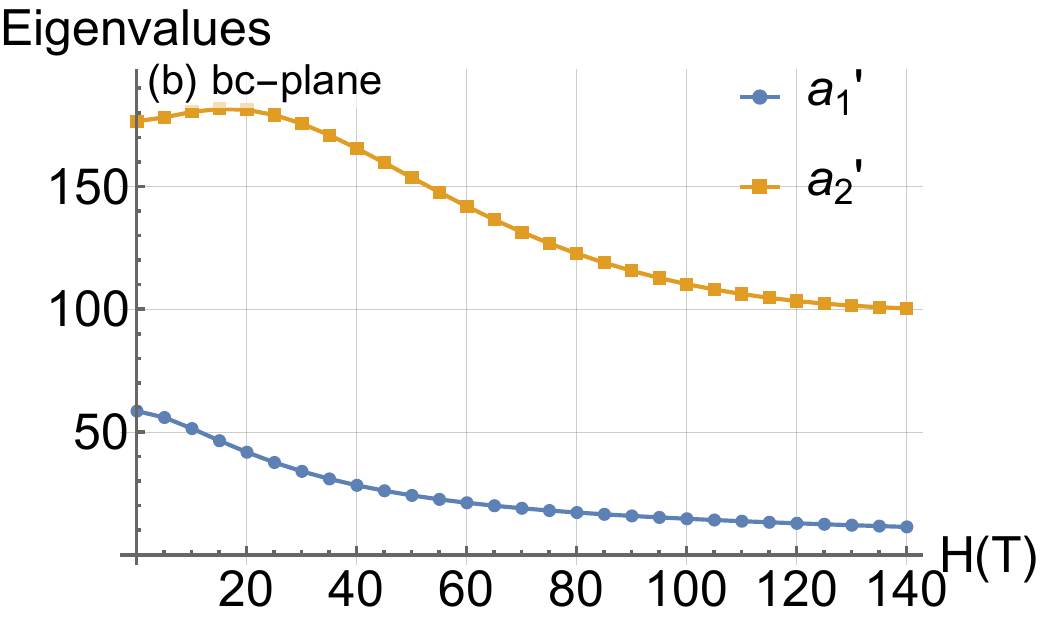}
\caption{Field dependence of the eigenvalues $(a_1, a_2)$ (panel a) and $(a_1', a_2')$ (panel b) of the Hessian matrices ${\bf A}$ and ${\bf A}'$, defined in (\ref{eq:HessianA}) and (\ref{eq:HessianAp}), which correspond to the $ac$- and $bc$-plane case, respectively. The data shown are taken at the representative angles $\theta_{ac}\!=\!81^\circ$ 
(measured from the $a$-axis) and $\theta_{bc}\!=\!45^\circ$  
(measured from the $b$-axis), for fields in the $ac$- and $bc$-plane, respectively, and for the coupling parameters $(J,K,\Gamma)=(0.4, -18, -10)$~meV and an isotropic $g$-tensor  with $g=2$.}
\label{fig:Eigenvalues}
\end{figure}

Figure~\ref{fig:Eigenvalues}\,(a) shows the two eigenvalues $(a_1, a_2)$ of the Hessian matrix ${\bf A}$ for a representative field direction in the $ac$ plane ($\theta_{ac}\!=\!0.45\pi$),  for which there is an intermediate phase. This graph shows that, as we decrease the field, the lowest eigenvalue $a_1$ of ${\bf A}$ vanishes at $H\!\simeq\!45.5$\,T, which is above the value of $H^\star$ at the same angle $\theta_{ac}$. Hence, for this angle, there is an intermediate phase that breaks the mirror symmetry ${\Theta}C_{2b}$. 
\\

\noindent{\bf $bc$-plane:}\\
For fields in the $bc$-plane, the PM phase has 
\be\label{eq:StatCond2bc}
\ra{1.25}
\begin{array}{ccc}
\overline{F}_b \neq 0\,,& \overline{F}_c\neq0\,,
&\overline{G}_a\neq0
\\
\overline{F}_a =
0\,,&\overline{G}_b=0\,,&\overline{G}_c=0\,,
\end{array}
\ee
which correspond to 
\be
\overline{\phi}=\pi/2\,,~~\overline{\alpha}=0\,.
\ee
The elements of the matrices $A$ and $B$ of Eq.~(\ref{eq:quadratic}), which we now call ${\bf A}'$ and ${\bf B}'$ (in order to differentiate them from the ones in $ac$-plane case), take the form
\be\label{eq:HessianAp}
\!\!\!
\ra{1.25}
\begin{array}{l}
A'_{\alpha\alpha} \!=\! 
\overline{G}_a^2 [
(g_c\!-\!g_a) \overline{F}_b^2
\!+\!(g_b\!-\!g_a)\overline{F}_c^2
]/\overline{{\bf F}}^2\!+\!\sqrt{2}\Gamma \overline{F}_b \overline{G}_a\,, 
\\
A'_{\phi\phi} \!=\! 
(f_a\!-\!f_b)\overline{F}_b^2
\!+\!(g_b\!-\!g_a)\overline{G}_a^2
\!+\!h_b\overline{F}_b
\!+\!4\sqrt{2}\Gamma \overline{F}_b \overline{G}_a
\,,
\\
A'_{\alpha\phi}\!=\!
[(g_a\!-\!g_b)\overline{G}^2_a\!-\!2\sqrt{2}\Gamma \overline{F}_b\overline{G}_a]\cos\overline{\theta}\,,
\end{array}
\ee
and
\be
\!\!\!
\ra{1.25}
\begin{array}{l}
B'_{\theta\theta}\!=\!(f_c\!-\!f_b)(\overline{F}_b^2\!-\!\overline{F}_c^2)\!+\! {\bf h}\cdot\overline{\bf F}\!+\! \sqrt{2}\Gamma \overline{F}_b\overline{G}_a\,,
\\
B'_{rr}\!=\!
[
(g_a\!-\!f_b)\overline{F}_b^2
\!+\!
(g_a\!-\!f_c)\overline{F}_c^2
][1-\frac{\overline{G}_a^2}{\overline{\bf F}^2}]
\!+\!
{\bf h}\cdot\overline{\bf F}\\
~~~~~~~~+ 
4\sqrt{2}\Gamma \overline{F}_b\overline{G}_a
\,,
\\
B'_{{\theta}r} \!=\!
[2(f_b\!-\!f_c) \overline{F}_b \overline{G}_a \!+\!\sqrt{2} \Gamma (\overline{\bf F}^2\!-\!\overline{G}_a^2) ]\cos\overline{\theta}
\\
~~~~~~~~
+\overline{G}_a (h_c \sin\overline{\theta}
\!-\!h_b \cos\overline{\theta})
\,,
\end{array}
\ee
where $\overline{\bf F}=(0, \overline{F}_b, \overline{F}_c)$ and ${\bf h}=(0, h_b, h_c)$.

Figure~\ref{fig:Eigenvalues}\,(b) shows the two eigenvalues $(a_1', a_2')$ of the Hessian matrix ${\bf A}'$ for a representative field direction in the $bc$ plane ($\theta_{bc}\!=\!\pi/4$). As shown in this graph, as we decrease the field, both eigenvalues remain positive for all field strengths, so there is no intermediate phase. The same is true for any other field direction in the $bc$ plane.

\begin{figure}[t!]
  \centering
\includegraphics[width=0.99\columnwidth,trim=0cm 0cm 0cm 0cm, clip=true]{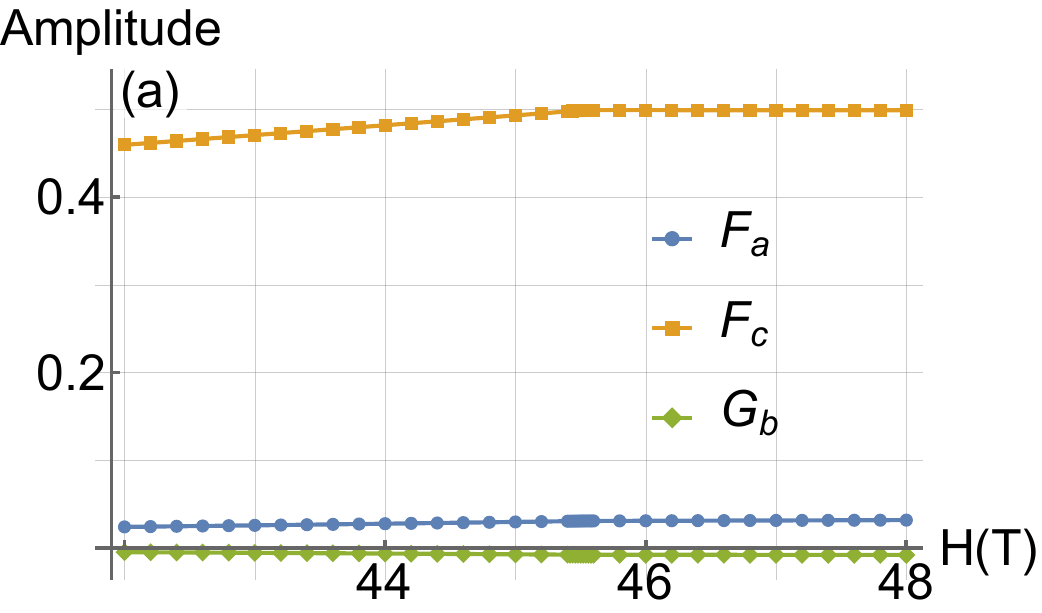}
\includegraphics[width=0.99\columnwidth,trim=0cm 0cm 0cm 0cm, clip=true]{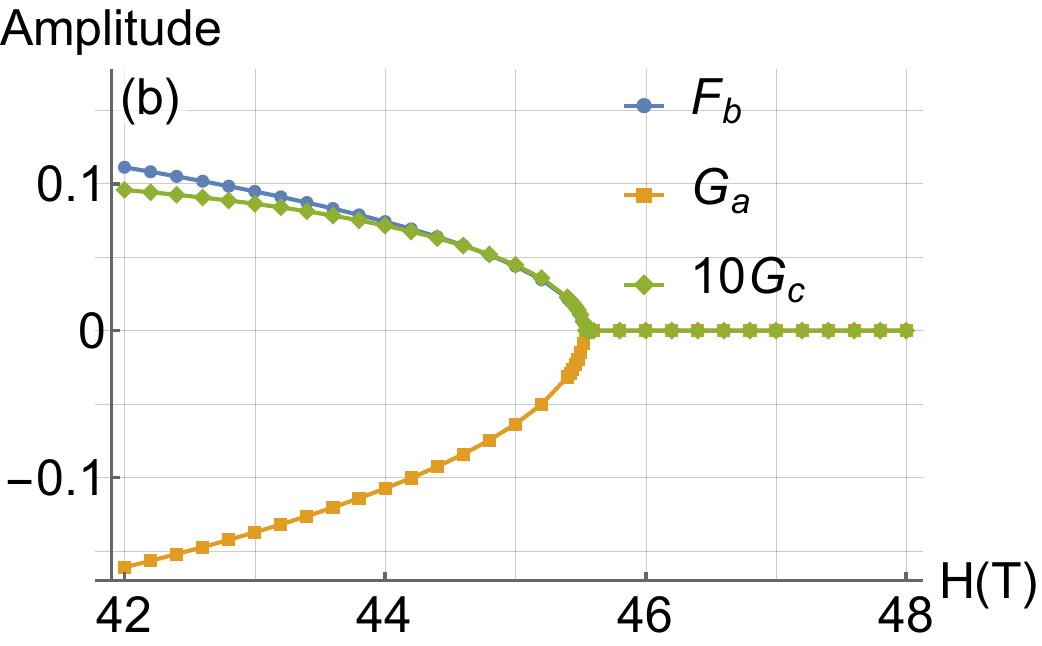}
\includegraphics[width=0.99\columnwidth,trim=0cm 0cm 0cm 0cm, clip=true]{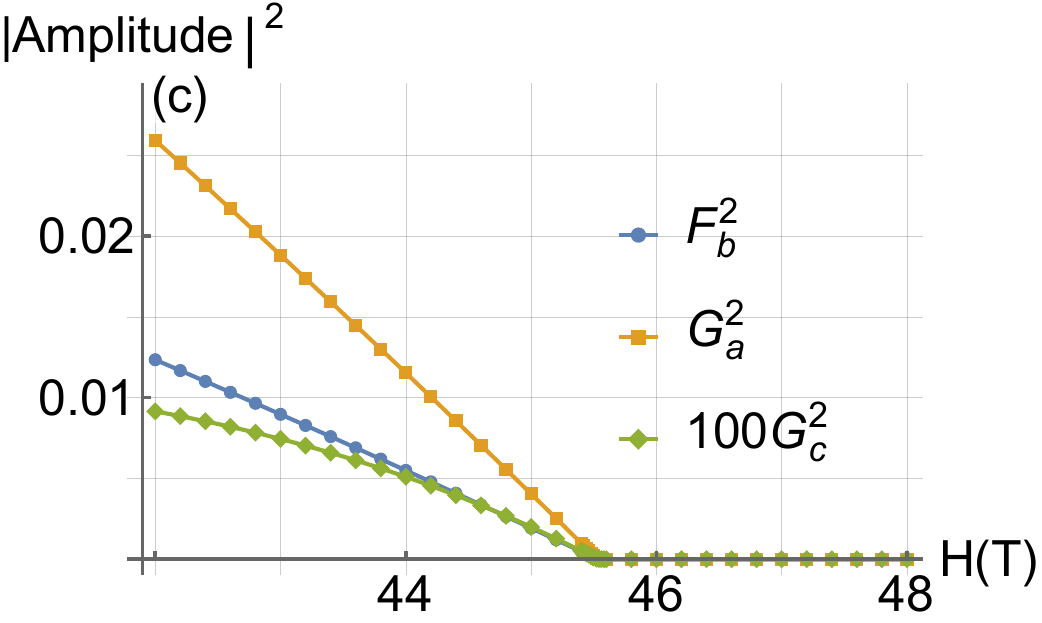}
\caption{Field dependence of the components of ${\bf F}$ and ${\bf G}$ around the critical field $H^{\star\star}$ for field in the $ac$-plane, with $\theta_{ac}\!=\!81^\circ$ 
(measured from the $a$-axis), $(J,K,\Gamma)=(-0.4,-18,-10)$\,meV, $g_{aa}\!=\!g_{bb}\!=\!g_{cc}\!=\!2$ and $g_{ab}\!=\!0$. (a) $(F_a, F_c, G_b)$ are driven by the field. (b) The order parameters $(F_b, G_a, G_c)$ appear spontaneously below $H_{ac}^{\star\star}$. (c) Below the transition, the order parameters scale as $\sqrt{H_{ac}^{\star\star}-H}$.}\label{fig:FieldDependence}
\end{figure}

\subsection{Order parameter field-dependence below $H_{ac}^{\star\star}$}\label{subsec:LandauAnalysis2}

Having established that an intermediate phase emerges only for magnetic fields in the $ac$-plane, we now examine how its order parameters develop below the critical field $H_{ac}^{\star\star}$. At the classical mean-field level, the  order parameters $F_b$, $G_a$ and $G_c$ of the intermediate phase found for fields in the $ac$ plane grow like 
\be\label{eq:OPdH}
F_b, G_a, G_c \propto \left(H_{ac}^{\star\star}-H\right)^{1/2}\,.
\ee 
To show this, we use the representation of ${\bf F}$ and ${\bf G}$ in terms of their components in the $abc$ frame and minimize the energy in Eqs.~(\ref{eq:EvsFG}) and (\ref{eq:HZ2}) subject to the constraints (\ref{eq:constraints}). To that end, we introduce two Lagrange multipliers $\mu_1$ and $\mu_2$, which enforce the constraints (\ref{eq:constraints}), and define the `free energy' 
\be\label{freeenergy}
\mc{F} = \mc{H}-\frac{\mu_1}{2}({\bf F}^2+{\bf G}^2-S^2)-\mu_2 {\bf F}\cdot{\bf G}\,.
\ee
Minimizing gives the stationary point conditions
\be\label{eq:StatCond1}
\ra{1.25}
\begin{array}{l}
(f_a-\mu_1) F_a - \sqrt{2}\Gamma G_b - \mu_2 G_a - h_a = 0\,,
\\
(f_b-\mu_1) F_b - \sqrt{2}\Gamma G_a - \mu_2 G_b - h_b= 0\,,
\\
(f_c-\mu_1) F_c - \mu_2 G_c - h_c = 0\,,
\\
(g_a-\mu_1) G_a - \sqrt{2}\Gamma F_b - \mu_2 F_a = 0\,,
\\
(g_b-\mu_1) G_b - \sqrt{2}\Gamma F_a - \mu_2 F_b = 0\,,
\\
(g_c-\mu_1) G_c - \mu_2 F_c = 0\,.
\end{array}
\ee
For $H\!\geq\!H^{\star\star}_{ac}$, the stationary point equations must give (\ref{eq:StatCond2}). Replacing into (\ref{eq:StatCond1}) gives  
\be\label{eq:StatCond2b}
\ra{1.25}
\begin{array}{l}
\overline{\mu}_2=0\,,
\\
\overline{F}_c = \overline{h}_c/ (f_c-\overline{\mu}_1)\,,
\\
\overline{F}_a = \overline{h}_a/ (f_a-\overline{\mu}_1-\frac{2\Gamma^2}{g_b-\overline{\mu}_1})\,,
\\
\overline{G}_b = \sqrt{2}\Gamma \overline{F}_a / (g_b-\overline{\mu}_1)\,,
\end{array}
\ee
where $\overline{h}_a=H_{ac}^{\star\star} \cos\theta_{ac}$, $\overline{h}_c=H_{ac}^{\star\star} \sin\theta_{ac}$. The value of $\mu_1$ at $H_{ac}^{\star\star}$, $\overline{\mu}_1$, can be found by solving
\be\label{eq:const1}
\overline{F}_a^2+\overline{F}_c^2+\overline{G}_b^2 = S^2\,.
\ee
The constraint $\overline{{\bf F}}\cdot\overline{{\bf G}}\!=\!0$ is automatically satisfied.

Incidentally, for fields along the $c$-axis, Eqs.~(\ref{eq:StatCond2b}) and (\ref{eq:const1}) give $\overline{F}_a\!=\!\overline{G}_a\!=\!0$ and $\overline{F}_c\!=\!S$ at $H\!\geq\!H_c^{\star\star}$. So, at zero temperature, and for fields along the $c$-axis, the system enters the {\it fully} polarised state above $H_c^{\star\star}$, as shown in Ref.~\cite{li-2020-uncon-magnet}.

We now consider fields just below the transition,
$H=H_{ac}^{\star\star}-\delta H$ (with $\delta H$ small enough). Since the transition is continuous, the new order parameters are expected to remain small. We therefore expand around the stationary solution at $H_{ac}^{\star\star}$ according to
\be\label{eq:Expand}
\ra{1.25}
\begin{array}{ccc}
F_a\!=\!\overline{F}_a\!+\!\delta F_a,
& 
F_c\!=\!\overline{F}_c\!+\!\delta F_c,
& 
G_b\!=\!\overline{G}_b\!+\!\delta G_b,
\\
F_b \!=\! \delta F_b,
&
G_a \!=\! \delta G_a,
&
G_c\!=\!\delta G_c .
\end{array}
\ee
The instability is associated with a single soft mode, whose amplitude we denote by $\psi$. Accordingly, we write
\be
\delta F_b = u \psi\,,~~~
\delta G_a = v \psi\,,
\ee
where $u$ and $v$ are some numerical coefficients  corresponding to the eigenvector of the Hessian matrix of the {\it effective} free energy (not analyzed here) of $F_b$ and $G_a$.  

The orthogonality constraint ${\bf F}\cdot{\bf G}=0$ implies that $G_c$ is not an independent order parameter. Instead, it is induced by the primary order parameters $F_b$ and $G_a$. Specifically,
\be
\!\!\delta G_c \!=\! -\frac{F_a G_a\!+\!F_b G_b}{F_c} \!=\! -\frac{v \overline{F}_a\!+\!u \overline{G}_b}{\overline{F}_c}\psi\!+\!\mc{O}(\psi^2)\,.
\ee 
The remaining constraint gives 
\be\label{eq:const2}
\sum_{\nu=a,b,c}(\overline{F}_\nu+\delta F_\nu)^2+(\overline{G}_\nu+\delta G_\nu)^2= S^2\,.
\ee
Subtracting (\ref{eq:const1}) from (\ref{eq:const2}) and keeping the leading terms in each of the deviations $\delta F_\nu$ and $\delta G_\nu$ gives
\be
\!\!2(\overline{F}_a\delta F_a \!+\!\overline{F}_c\delta F_c\!+\!\overline{G}_b\delta G_b)\!+\!\delta F_b^2\!+\!\delta G_a^2\!+\!\delta G_c^2\!=\!0.
\ee
This tells us that $\delta F_a$, $\delta F_c$ and $\delta G_b$ scale at least as $\psi^2$. 
Furthermore, replacing (\ref{eq:Expand}) into  (\ref{eq:StatCond1}), we learn that $\delta \mu_2 \propto \psi$, $\delta \mu_1 \propto \psi^2$, and also that 
\be
\psi \propto \sqrt{\delta H}\,,
\ee
which explains (\ref{eq:OPdH}). 

The above can also be seen  in Fig.~\ref{fig:FieldDependence}, which shows the field dependence of all components of ${\bf F}$ and ${\bf G}$, for a representative field orientation in the $ac$-plane ($\theta_{ac}\!=\!0.45\pi$). Panel (a) shows that the changes in $(F_a, F_c, G_b)$ below $H_{ac}^{\star\star}$ scale linearly with $\delta H$, whereas panel (c) shows that the order parameters $(F_b, G_a, G_c)$ are proportional to $\sqrt{\delta H}$.

\subsection{Torque and magnetotropic coefficient}
In this subsection, we provide auxiliary information for the dependence of the torque and the magnetotropic coefficient on the components of ${\bf F}$ and ${\bf G}$. 
In the following, ${\bf m}$ denotes the magnetic moment per site and $\bs{\tau}={\bf m} \times {\bf H}$ is the torque per site.

\subsubsection{Fields in the $bc$-plane}
Here the torque is along the $a$-axis, and is given by~\cite{li-2020-reent-incom}
\be
\tau_a / H = m_b \sin\theta_{bc} - m_c \cos\theta_{bc}\,,
\ee
with 
\be
m_b = \mu_B
(g_{bb} \overline{F}_b \!-\! g_{ab} \overline{G}_a)\,,~~
m_c = g_{cc}\mu_B\overline{F}_c\,. 
\ee
The magnetotropic coefficient $k_{bc}\!=\!\frac{d\tau_a}{d\theta_{bc}}$ takes the form
\be
\frac{k_{bc}}{\mu_B H} \!=\!
(m_b-m_c') \cos\theta_{bc}
+ (m_b'+m_c) \sin\theta_{bc}\,, 
\ee
where $m_b'$ and $m_c'$ stand for $\frac{dm_b}{d\theta_{bc}}$ and $\frac{dm_c}{d\theta_{bc}}$, respectively. 
Hence, $\tau_a$ and $k_{bc}$ depend on the field-driven components $\overline{F}_b$, $\overline{F}_c$ and $\overline{G}_a$.

\subsubsection{Fields in the $ab$-plane}
Here the torque is along the $c$-axis, and is given by~\cite{li-2020-reent-incom}
\be
\tau_c / H = 
m_a \cos\theta_{ab} - m_b \sin\theta_{ab}\,,
\ee
with
\be
m_a \!=\! \mu_B
(g_{aa} \overline{F}_a \!-\! g_{ab} \overline{G}_b),~
m_b\!=\!\mu_B (g_{bb}\overline{F}_b\!-\!g_{ab}\overline{G}_a).
\ee
The magnetotropic coefficient $k_{ab}\!=\!\frac{d\tau_c}{d\theta_{ab}}$ takes the form
\bea
\frac{k_{ab}}{\mu_B H} \!=\!
-(m_a+m_b') \sin\theta_{ab} 
+(m_a'- m_b) \cos\theta_{ab}\,,
\eea
where $m_a'$ and $m_b'$ stand for $\frac{dm_a}{d\theta_{ab}}$ and $\frac{dm_b}{d\theta_{ab}}$, respectively. 
Hence, $\tau_c$ and $k_{ab}$ depend on the field-driven terms $\overline{F}_a$, $\overline{F}_b$,  $\overline{G}_a$ and $\overline{G}_b$.

\subsubsection{Fields in the $ac$-plane}
Here, all components of the torque in the $abc$-frame are nonzero. The $b$-component of the torque, which is relevant for the magnetotropic coefficient,  is given by~\cite{li-2020-reent-incom}
\be
\tau_b/H = m_c \cos\theta_{ac}-m_a\sin\theta_{ac}\,,
\ee
where
\be
m_a\!=\!\mu_B(g_{aa}\overline{F}_a-g_{ab} \overline{G}_b),~
m_c \!=\! g_{cc} \mu_B \overline{F}_c.
\ee
The magnetotropic coefficient $k_{ac}\!=\!\frac{d\tau_b}{d\theta_{ac}}$ takes the form
\bea
\frac{k_{ac}}{\mu_B H} \!=\!
(m_c-m_a') \sin\theta_{ab} 
+(m_c'- m_a) \cos\theta_{ab}\,,
\eea
where $m_a'$ and $m_c'$ stand for $\frac{dm_a}{d\theta_{ac}}$ and $\frac{dm_c}{d\theta_{ac}}$, respectively. 
Hence, $\tau_b$ and $k_{ac}$ depend on the field-driven terms $\overline{F}_a$, $\overline{F}_c$ and $\overline{G}_b$.

\bibliography{references_final}